# The Role of Feedback in Two-way Secure Communications

Xiang He    Aylin Yener

Wireless Communications and Networking Laboratory

Electrical Engineering Department

The Pennsylvania State University, University Park, PA 16802

*xxh119@psu.edu*    *yener@ee.psu.edu*

November 20, 2009

**Abstract**

Most practical communication links are bi-directional. In these models, since the source node also receives signals, its encoder has the option of computing its output based on the signals it received in the past. On the other hand, from a practical point of view, it would also be desirable to identify the cases where such an encoder design may not improve communication rates. This question is particularly interesting for the case where the transmitted messages and the feedback signals are subject to eavesdropping. In this work, we investigate the question of how much impact the feedback has on the secrecy capacity by studying two fundamental models. First, we consider the Gaussian two-way wiretap channel and derive an outer bound for its secrecy capacity region. We show that the secrecy rate loss can be unbounded when feedback signals are not utilized except for a special case we identify, and thus conclude that utilizing feedback can be highly beneficial in general. Second, we consider a half-duplex Gaussian two-way relay channel where the relay node is also an eavesdropper, and find that the impact of feedback is less pronounced compared to the previous scenario. Specifically, the loss in secrecy rate, when ignoring the feedback, is quantified to be less than 0.5 bit per channel use when the relay power goes to infinity. This achievable rate region is obtained with simple time sharing along with cooperative jamming, which, with its simplicity and near optimum performance, is a viable alternative to an encoder that utilizes feedback signals.

This work was presented in part at the 42nd Asilomar Conference on Signals, System and Computers, October 2008. This work is supported in part by the National Science Foundation via Grants CCR-0237727, CCF-051483, CNS-0716325, and the DARPA ITMANET Program via Grant W911NF-07-1-0028.

## I. Introduction

Most communication links are bi-directional, where the backward channel can carry information and/or provides some form of feedback. For example, in ARQ schemes, the backward channel provides the acknowledgment of receipt of the packets. In peer-to-peer networks, information is communicated in both directions. The impact of the existence of bi-directionality on the channel capacity has been considered extensively up to date. Shannon proposed the two-way channel model in [1] where communication took place in both directions, and derived the inner bound and the outer bound on its capacity region. These bounds were shown to match for the full-duplex Gaussian two-way channel in [2]. An interesting implication of this result is that the signals received in the past, i.e., the feedback signals, is not needed for encoding to achieve the capacity region for this model. Though this feature is desirable in practice for simpler encoder design, it is also known that this approach is suboptimal in general, which was proved in [3] for a two-way channel where the two nodes share a common output from the channel.

In secure communication, the question of whether feedback signals should be used for encoding has been studied in several special scenarios. Shannon showed that a completely secure backward channel can be used to send a “one-time pad” to increase the secrecy capacity of the forward channel [4]. In [5], it was proved that such a strategy, where the source node decodes the key from the destination, is optimal for a degraded wiretap channel with a secure rate limited noiseless feedback link. Another achievable scheme, which does not require decoding of the feedback, was first proposed in [6] in the setting of secret key generation and later in [7]. The scheme proves even if the forward channel and backward channel each has zero secrecy capacity and hence sending key back is not possible, a positive secrecy rate can still be achieved when these two channels are used together. This is done by combining multiple channel uses and designing codes for the resulting equivalent broadcast channel in which the eavesdropper is eventually put at its disadvantage because of its lack of side information. Reference [8] combines this scheme with the key strategy in [4] and shows a higher secrecy rate is achievable for the model in [7].

In [5], [7], [8], the destination has the freedom to design the feedback signals. References [8], [9] also considered the scenario where the destination was restricted to sending its observation of the channel output, and hence could not manipulate the feedback signal to its advantage. It was shown that feedback also helped to achieve a higher secrecy rate in this case.

One feature that is common to the coding schemes in [5], [7], [8] is that the eavesdropper always receives two separate sets of received signals: one from the forward channel and a second set of signals from the backward channel if it is not secure. While this is more inline with the conventional information theoretic models with feedback [10, Section 7.12] [11], letting the eavesdropper receiving the signals of the forward and the backward channel separately might inadvertently give the eavesdropper an advantage, as compared to superimposing them together. Specifically, when the eavesdropper receives the sum of the outputs from the forward and the backward channel, introducing artificial noise into the backward channel at the time when the forward channel is in use can interfere the eavesdropper's observation of the forward channel and hence reduce its recognizance of the message being transmitted on it. This so-called "cooperative jamming" scheme has been shown to improve secrecy rates in a Gaussian two-way channel with an external eavesdropper [12]. Yet in reference [12], the source node does not take advantage of the signals it received from the backward channel when encoding its transmission signals. The question remains, therefore, in such a "cooperative jamming" scheme, whether the achievable rates can be improved by utilizing these signals.

In this paper, we consider the wireless communication scenario where the eavesdropper observes the sum of the outputs of the forward and the backward channel, and hence the legitimate nodes in the network can potentially utilize both feedback signals and cooperative jamming to protect the confidential message. We focus on two models where both techniques are potentially useful: (i) a class of Gaussian full-duplex two-way wiretap channels, and (ii) a Gaussian half-duplex two-way relay channel with an untrusted relay.

For the first model, we derive a computable outer bound to its secrecy capacity region. We then compare it to the achievable rates when the feedback is ignored at both nodes. Interestingly, when the ratio of the power constraint of the two legitimate nodes is fixed and the channel is fully connected with independent link noise, the gap between the achieved secrecy rate and the outer bound is bounded by a constant, which only depends on the channel gains.

On the other hand, when the ratio of the power constraints is not fixed, we show that ignoring feedback signals leads to unbounded loss in the secrecy rate when the power increases. The loss is measured as the gap between the achievable rate when the feedback is used and the upper bound when the feedback is not used, hence is not caused by the potential sub-optimality of the achievable scheme. This result shows that utilizing the feedback for encoding at the legitimate nodes is highly beneficial for this model in general.

In the second model, we consider the case where the eavesdropper is part of the network rather than being external to it. In this model, two nodes wish to exchange information via a relay node from whom the information needs to be kept secret. Here the relay node is "honest but curious" [13], in that it will faithfully carry out designated relaying scheme, but is not trusted to decode the message it is relaying. This kind of setting was first considered in [14] for the three node relay channel and later thoroughly studied in [15] and [16]. Later, in [17], we considered a restricted version of the model in this work, by studying the case when the feedback signals were not used at the source or the destination for encoding purposes. In this paper, we identify one case where doing so will not incur much loss in secrecy rate. More specifically, we will prove if the power of the relay goes to $\infty$, then the loss in the secrecy rates caused by ignoring the feedback is bounded by $0.5$ bit per channel use. Interestingly, a simple TDMA scheme with cooperative jamming yields the achievable rate.

The channel models in this work are closely related the the channel-type model in secret key generation literature; see [6], [18]–[21] for example. The major difference from these works is that our model accepts two inputs, one from the source, the other from the destination. The eavesdropper observes a noisy superposition of these two inputs. This is more complicated than the channel-type model where the noisy part of the channel is a wiretap channel which only accepts one input from the source node, and any input from the destination can only be transmitted over a noiseless public discussion link which is orthogonal to the wiretap channel. Recently, reference [22] has considered a channel-type secret key generation model where the channel component in the model accepts inputs from multiple nodes. Yet, these nodes only receive from the noiseless public discussion link [22, Section II], which is a fundamentally different model from those considered in this work.

The rest part of the paper is organized as follows: In Section II, we describe the two models considered in this work. Section III focuses on the Gaussian two-way wiretap channel. Section IV focuses on the two-way relay channel with an untrusted relay. Section V presents some alternative proofs to some results in previous sections. Section VI concludes the paper.

Throughout the paper the notation $C(x)$ is defined as $C(x) = \frac{1}{2}\log_2(1+x)$. Also $x_i$ denotes the $i$th component of vector $x$, while $x^i$ denotes $\{x_1, ...x_i\}$. $\mathcal{N}(0, \sigma^2)$ denotes a zero mean Gaussian

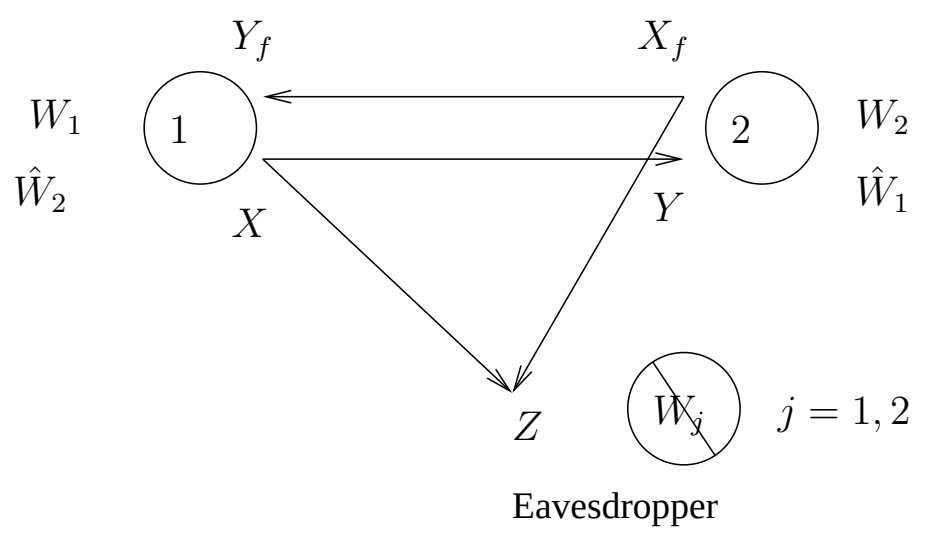

Fig. 1. Two-way wiretap channel

distribution with variance $\sigma^2$.

## II. Channel Models

In this section, we describe the two channel models considered in this work. Both models involve information exchange between two nodes: Node $1$ and Node $2$. Node $1$ wants to send a message $W_1$ to Node $2$. Node $2$ wants to send a message $W_2$ to Node $1$. Both messages must be kept secret from the eavesdropper. The encoding functions used at the two nodes are allowed to be stochastic. Without loss of generality, we use $M_j$ to model the local randomness in the encoding function used by Node $j$, $j = 1, 2$.

### *A. The Two-Way Wiretap Channel*

The first model we consider in this work is a two-way wiretap channel model. The channel model is shown in Figure 1. The channel description is given by

$$\Pr(Y, Y_f, Z|X, X_f) = \Pr(Z|X, X_f) \Pr(Y|X, X_f, Z) \Pr(Y_f|X_f, X, Z) \tag{1}$$

From (1), we observe

$$Y_f - \{X_f, X, Z\} - Y \tag{2}$$

is a Markov chain.

At each channel use, Node $1$ and Node $2$ transmit simultaneously. At the $i$th channel use, the encoding function of Node $1$ is defined as:

$$X_i = f_i(Y_f^{i-1}, W_1, M_1) \tag{3}$$

The encoding function of Node 2 is defined as

$$X_{f,i} = g_i(Y^{i-1}, W_2, M_2) \tag{4}$$

Note that with the introduction of $M_j, j = 1, 2$, we can define $f_i, g_i$ as deterministic encoders. Also note that another way to define $f_i$ is $X_i = f_i(X^{i-1}, Y_f^{i-1}, M_1)$. It is easy to see that this definition is equivalent to the definition given in (3).

Let $n$ be the total number of channel uses. Node 2 must decode $W_1$ reliably from $X_f^n, Y^n, M_2, W_2$. Node 1 must decode $W_2$ reliably from $Y_f^n, X^n, M_1, W_1$. Let the decoding results be $\hat{W}_1$ and $\hat{W}_2$ respectively. Then we require

$$\lim_{n\to\infty} \Pr(W_j \neq \hat{W}_j) = 0, \quad j = 1, 2 \tag{5}$$

Hence, from Fano's inequality [10], we have

$$H(W_1|X_f^n, Y^n, M_2, W_2) < n\varepsilon_1 \tag{6}$$

$$H(W_2|Y_f^n, X^n, M_1, W_1) < n\varepsilon_2 \tag{7}$$

where $\varepsilon_j > 0$ and $\lim_{n\to\infty} \varepsilon_j = 0$, $j = 1, 2$.

In addition, both messages must be kept secret from the eavesdropper. Hence

$$I(W_1, W_2; Z^n) < n\varepsilon_3 \tag{8}$$

where $\varepsilon_3 > 0$ and $\lim_{n\to\infty} \varepsilon_3 = 0$.

Define $R_j, j = 1, 2$ as:

$$R_j = \lim_{n\to\infty} \frac{1}{n} H(W_j), \quad j = 1, 2 \tag{9}$$

The secrecy rate region is defined as all rate pairs $\{R_1, R_2\}$ for which (5) and (8) holds.

The Gaussian case of the two-way wiretap channel model was first proposed in [12] and is shown in Figure 2. Formally, the channel is described as:

$$Y_f = X_f + N_3 + \sqrt{\alpha} X \tag{10}$$

$$Y = X + N_1 + \sqrt{\beta} X_f \tag{11}$$

$$Z = \sqrt{h_1} X + \sqrt{h_2} X_f + N_2 \tag{12}$$

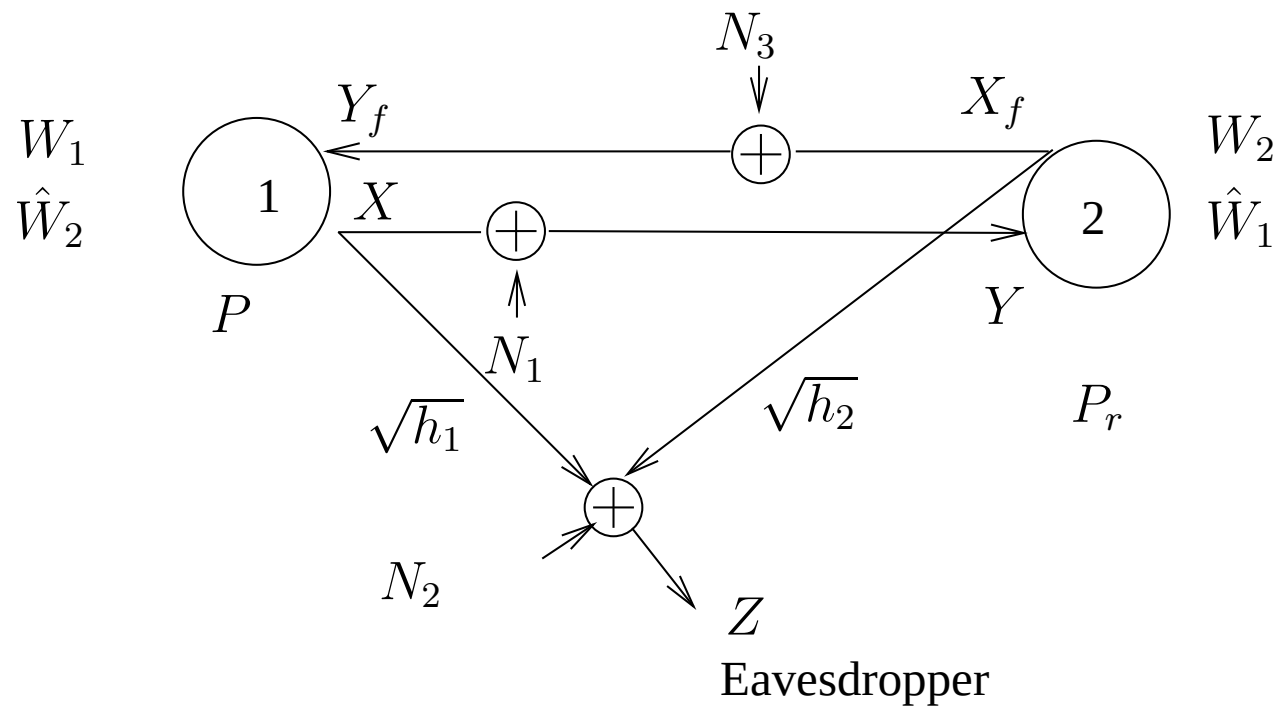

Fig. 2. The Gaussian Two-way Wiretap Channel

where $\sqrt{\alpha}, \sqrt{\beta}, \sqrt{h_1}, \sqrt{h_2}$ are channel gains. $N_i, i = 1, 2, 3$ are Gaussian random variables with zero mean and unit variance, representing the channel noise. We assume that given $N_2$, $N_1$ is independent from $N_3$:

$$p(N_1, N_2, N_3) = p(N_2)p(N_1|N_2)p(N_3|N_2) \tag{13}$$

We use $\rho$ to denote the correlation factor between $N_1$ and $N_2$. $\eta$ denotes the correlation factor between $N_2$ and $N_3$. Obviously, $-1 \leq \rho \leq 1$, and $-1 \leq \eta \leq 1$.

From (1) and (13), we readily see this channel belongs to the class of channels described by (1) and shown in Figure 1.

Observe that the terms $\sqrt{\alpha}X$ and $\sqrt{\beta}X_f$ are not shown in Figure 2. This is because each node knows its own transmitted signal and $\sqrt{\alpha}, \sqrt{\beta}, \sqrt{h_1}, \sqrt{h_2}$, and can always subtract the interference caused by its own transmitted signals. Hence we can remove $\sqrt{\alpha}X$ and $\sqrt{\beta}X_f$ from (10) and (11). The channel is hence equivalent to

$$Y_f = X_f + N_3 \tag{14}$$

$$Y = X + N_1 \tag{15}$$

$$Z = \sqrt{h_1}X + \sqrt{h_2}X_f + N_2 \tag{16}$$

In the sequel we shall focus on this equivalent model instead.

Let the power constraint of Node 1 be $P$. Let the power constraint of Node 2 be $P_r$.

$$\frac{1}{n}\sum_{k=1}^{n} E\left[X_k^2\right] \leq P \tag{17}$$

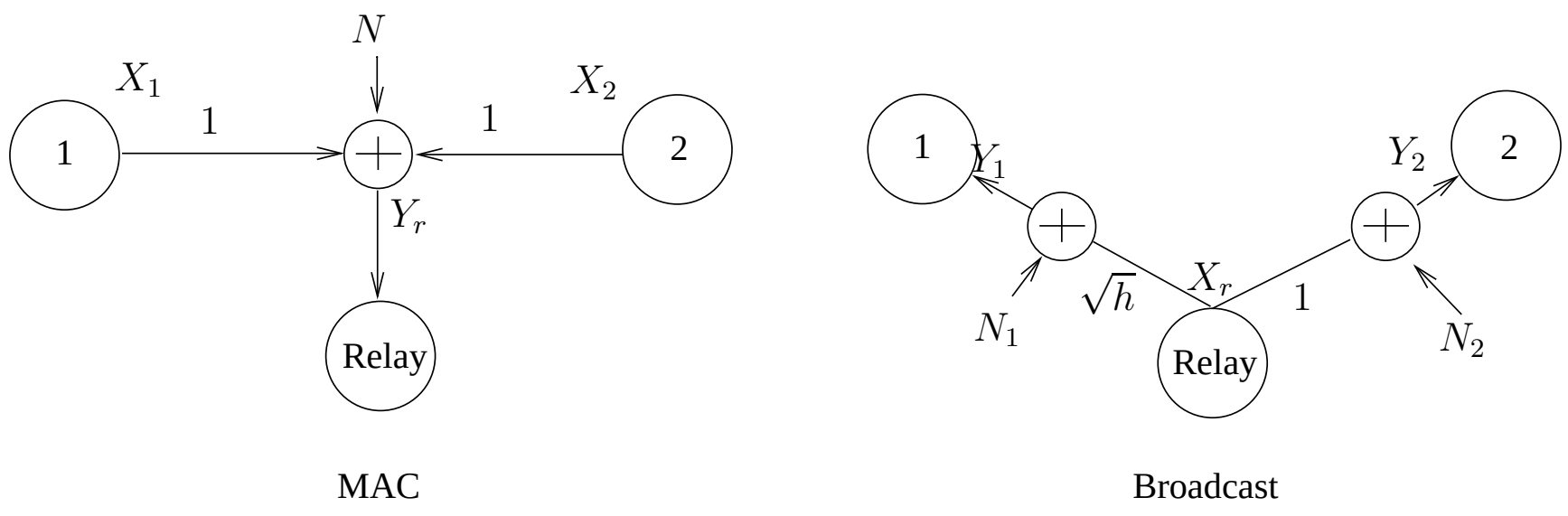

Fig. 3. The Gaussian two-way half-duplex relay channel with an untrusted relay

$$\frac{1}{n}\sum_{k=1}^{n} E\left[X_{f,k}^2\right] \leq P_r \tag{18}$$

*Remark 1:* When $Y_f$ is a constant, or, the feedback is ignored by Node 1, the model reduces to the relay channel with a confidential message to the relay, which was considered in references [16], [23], [24]. □

## *B. Two-Way Relay Channel with an Untrusted Relay*

The second model we consider in this work is the Gaussian two-way relay channel with an untrusted relay node. The channel model is shown in Figure 3. At any time slot, the channel either behaves as a MAC channel, shown on the left, or as a broadcast channel, shown on the right. After normalizing the channel gains, the MAC channel can be expressed as:

$$Y_r = X_1 + X_2 + N \tag{19}$$

The broadcast channel can be expressed as:

$$Y_1 = \sqrt{h}X_r + N_1 \tag{20}$$

$$Y_2 = X_r + N_2 \tag{21}$$

where $\sqrt{h}$ is the channel gain, $h \neq 0$. $N$, $N_1$, $N_2$ are independent zero mean Gaussian random variables with unit variance.

We assume Node 1 and Node 2 transmit simultaneously during the MAC mode. $X_{j,i}, j = 1, 2$ denote the signals transmitted by Node $j$ during the $i$th channel use such that the channel is in MAC mode. $i \geq 1$. We use $\phi_i$ to denote the number of channel uses that the channel was in the broadcast mode before this channel use. The notation $X_j^i$ denotes the set of signals: $\{X_{j,k}, k = 1...i\}$.

Similarly $X_{r,i}$ denotes the signal transmitted by the relay node during the $i$th channel use that the channel is in broadcast mode. $i \geq 1$. We use $\psi_i$ to denote the number of channel uses that the channel was in the MAC mode before this channel use.

$Y_{1,i}, Y_{2,i}, Y_{r,i}$ are received signals defined in the same fashion.

The channel switches between the MAC mode and the broadcast mode according to a globally known schedule. We assume the schedule is independent from the local randomness at each node, the messages and the channel noise. The first mode is assumed to be the MAC mode. The case where the first mode is a broadcast mode can be viewed as a special case of invoking the MAC mode first by transmitting nothing during the first MAC mode. The rate loss caused by the wasted channel use is negligible as the number of channel uses goes to $\infty$.

Suppose the MAC mode is activated for $n$ channel uses. The broadcast mode is activated for $m$ channel uses. Hence the communication spans over $n+m$ channel uses. It should be noted that, in general, neither the $n$ channel uses of the MAC mode, nor the $m$ channel uses of the broadcast mode have to be consecutive. We assume the schedule is stable, in the sense that the following limit exists:

$$\alpha = \lim_{n+m \to \infty} \frac{n}{m+n} \tag{22}$$

For a given $\alpha$, we use $\{T(\alpha)\}$ to denote a sequence of schedules with increasing total number of channel uses $n+m$ such that (22) holds, and $\alpha$ is the limit of the time sharing factor of the MAC mode in the schedule $T(\alpha)$ as $n+m \to \infty$.

The average power constraints for the source, the jammer and the relay can be expressed as:

$$\frac{1}{m+n} \sum_{k=1}^{n} E\left[X_{i,k}^2\right] \leq \bar{P}_i, \quad i = 1, 2, \tag{23}$$

$$\frac{1}{m+n} \sum_{k=1}^{m} E\left[X_{r,k}^2\right] \leq \bar{P}_r \tag{24}$$

For the purpose of completeness, we also introduce the notation $P_i, i = 1, 2$ to denote the average power of Node $i$ during the MAC mode. Since these two nodes are only transmitting during the MAC model, $P_i$ and $\bar{P}_i$ are related as

$$P_i = \bar{P}_i / \alpha, \quad i = 1, 2 \tag{25}$$

Similarly, we use $P_r$ to denote the average power of the relay node during the broadcast mode. Since the relay node only transmits during the broadcast mode, $P_r$ is related to $\bar{P}_r$ as follows:

$$P_r = \bar{P}_r/(1-\alpha) \tag{26}$$

For the $i$th channel use in which the channel operates in the MAC mode, the encoding functions at Node 1, $f_{1,i}$, is defined as:

$$X_{1,i} = f_{1,i}(Y_1^{\phi_i}, W_1, M_1) \tag{27}$$

Similarly, the encoding functions at Node 2, $f_{2,i}$, is defined as:

$$X_{2,i} = f_{2,i}(Y_2^{\phi_i}, W_2, M_2) \tag{28}$$

Note that $f_{1,i}, f_{2,i}$ are deterministic functions, and we use $M_r$ to model the local randomness at the relay. For the $i$th channel use in which the channel operates in broadcast mode, the encoding function of the relay node, $g_i$, is defined as:

$$X_{r,i} = g_i(Y_r^{\psi_i}, M_r) \tag{29}$$

where $g_i$ is a deterministic function.

The eavesdropper knows $Y_r^n, X_r^m, M_r$. Therefore, the secrecy constraint is expressed as

$$\lim_{m+n\to\infty} \frac{1}{m+n} H(W_1, W_2|Y_r^n, X_r^m, M_r) = \lim_{m+n\to\infty} \frac{1}{m+n} H(W_1, W_2) \tag{30}$$

Since $W - \{X_r^m, Y_r^n\} - M_r$ is a Markov chain, we have

$$\lim_{m+n\to\infty} \frac{1}{m+n} H(W_1, W_2|Y_r^n, X_r^m, M_r) = \lim_{m+n\to\infty} \frac{1}{m+n} H(W_1, W_2|Y_r^n, X_r^m) \tag{31}$$

Therefore, the secrecy constraint can be expressed as

$$\lim_{m+n\to\infty} \frac{1}{m+n} H(W_1, W_2|Y_r^n, X_r^m) = \lim_{m+n\to\infty} \frac{1}{m+n} H(W_1, W_2) \tag{32}$$

Let $\hat{W}_j, j = 1, 2$ be the decoding result computed by the intended receiver of $W_j, j = 1, 2$. Then the reliable communication requirement is expressed as

$$\lim_{m+n\to\infty} \Pr(\hat{W}_j \neq W_j) = 0, j = 1, 2 \tag{33}$$

Define $R_1, R_2$ as

$$R_j = \lim_{m+n\to\infty} \frac{1}{n+m} H(W_j), j = 1, 2 \tag{34}$$

The secrecy capacity region is defined as the union of all rate pairs $(R_1, R_2)$ such that there is an $\alpha$, a sequence of schedule $\{T(\alpha)\}$ and a choice of encoding function for which (32) and (33) are satisfied.

*Remark 2:* In general,

$$\lim_{n+m\to\infty} \frac{1}{n+m} H\left(W|X_r^m, Y_r^n\right) \neq \lim_{n+m\to\infty} \frac{1}{n+m} H\left(W|Y_r^n\right) \tag{35}$$

This can be proved by a counterexample: Consider the communication protocol:

1) First the relay node randomly generates and broadcasts a key via $X_r$ to Node $1$ and Node $2$ using a channel code.
2) Node $1$ uses the key as a one-time pad [4] to encrypt its confidential message $W$ and sends it to the relay using a channel code. The other nodes remain silent.
3) The relay decodes the codeword sent by Node $1$ and encodes and forwards it to the destination.
4) The destination recovers the codeword sent by Node $1$ by decoding the signals from the relay. It then decrypts it with the key it received in step 1 and recovers $W$.

Since the one-time pad is a perfectly secure cipher [4], for this communication protocol, we have:

$$H\left(W\right) = H\left(W|Y_r^n\right) \tag{36}$$

However, since the key is determined by $X_r^m$, given the key, $W$ is uniquely determined by $Y_r^n$. Therefore, we have

$$H\left(W|X_r^m, Y_r^n\right) = 0 \neq H(W|Y_r^n) \tag{37}$$

□

## III. Feedback in the Two-way Wiretap Channel

### A. Improvement on the Known Achievable Secrecy Rate: A Motivating Example

For the two-way wiretap channel, reference [12] derived an achievable rate using Gaussian codebooks. However, in this scheme, the signal $Y_f$ received by Node $1$ is not used to compute the signal $X$ transmitted by Node $1$. Likewise, the signal $Y$ received by Node $2$ is not used to compute the signal $X_f$ transmitted by Node $2$. We next show that this scheme can be improved upon with respect to the achievable secrecy rate. To show this, it is sufficient to show that a larger $R_1$ is achievable for Node 1 for a set of channel gains. In the following, we provide such an example.

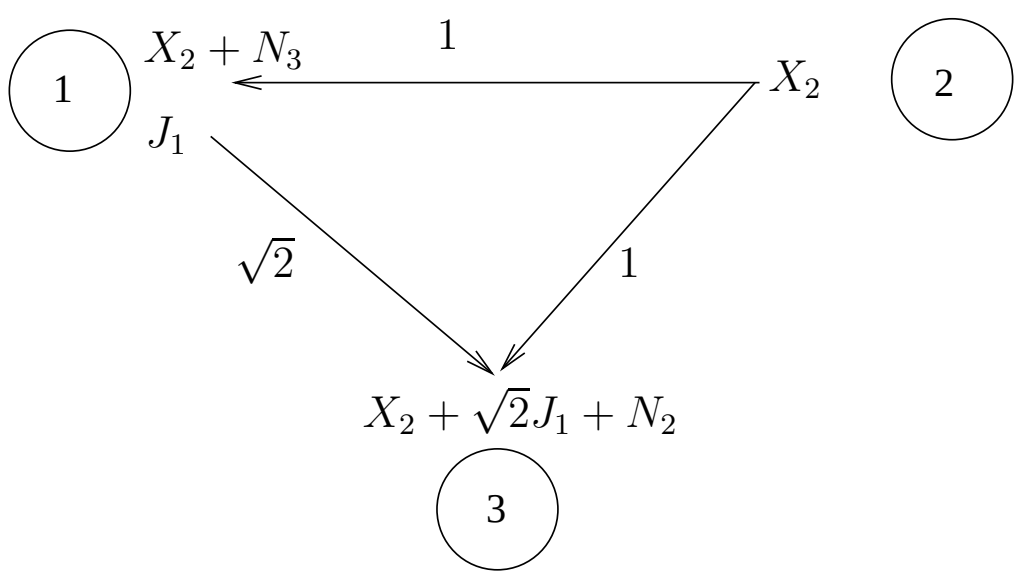

Fig. 4. Odd Step

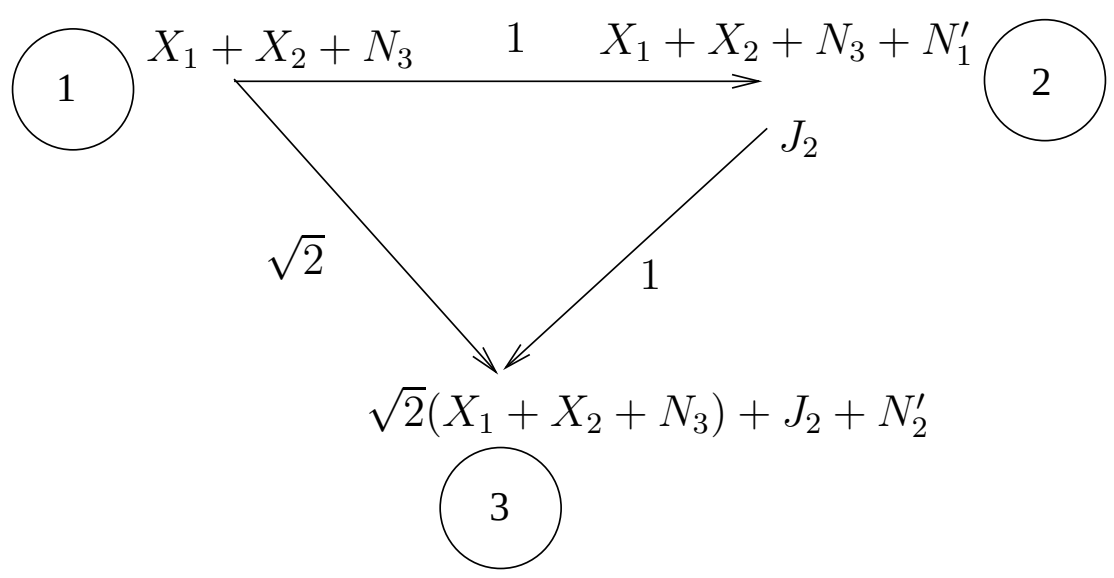

Fig. 5. Even Step

We assume $\rho = 0, \eta = 0$, which means $N_1, N_2, N_3$ are all independent, which was the setting considered by [12]. The largest rate for Node 1 achievable with the scheme of [12] is given by:

$$R_1 = [C(P) - C\left(\frac{h_1 P}{h_2 P_r + 1}\right)]^+ \tag{38}$$

which is achieved by letting Node 2 transmit an i.i.d. Gaussian sequence with variance $P_r$. When $\frac{h_1}{h_2 P_r + 1} \geq 1$, we observe from (38) that the secrecy rate is always 0. Below, we choose $P = 3, P_r = 1, \sqrt{h_1} = \sqrt{2}, \sqrt{h_2} = 1$ such that this condition is fulfilled and prove a positive secrecy rate is achievable with our scheme.

The coding scheme we use is similar to that of [6]. It is composed of one channel use described in Figure 4, followed by one channel use described in Figure 5. In an odd step, Node 1 sends a signal denoted by $J_1$ and Node 2 sends a signal denoted by $X_2$. After this step, Node 1 adds its received signal $X_2 + N_3$ to a new signal $X_1$ and transmits it in the following even step. At the same time, Node $2$ sends a signal denoted by $J_2$. We use the notation $N_i$ to denote the channel noise in the odd step and $N_i'$ to denote the channel noise in the even step.

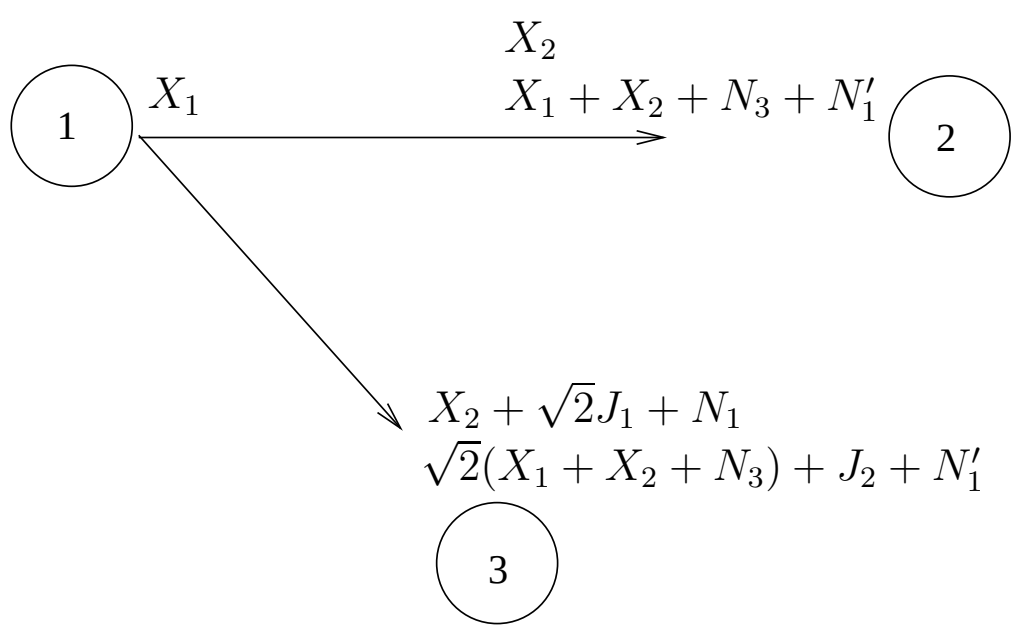

Fig. 6. The Equivalent Channel

Combining these two steps, we obtain an equivalent memoryless channel shown in Figure 6. The achievable secrecy rate for this channel is given by [25]:

$$[I\left(X_1;Y\right) - I\left(X_1;Y_{e,1},Y_{e,2}\right)]^+ \tag{39}$$

where

$$Y = X_1 + N_3 + N_1' \tag{40}$$

$$Y_{e,1} = X_2 + \sqrt{2}J_1 + N_2 \tag{41}$$

$$Y_{e,2} = \sqrt{2}\left(X_1 + X_2 + N_3\right) + J_2 + N_2' \tag{42}$$

We then choose $X_1, X_2, J_1, J_2$ as zero mean independent Gaussian random variables with unit variance. From Figures 4 and 5, this choice satisfies the average power constraints. Evaluating (39) for this distribution, we get

$$C\left(\frac{1}{2}\right) - C\left(\frac{a^2}{2a^2+2-\frac{a^2}{a^2+2}}\right) > 0 \tag{43}$$

where $a = \sqrt{2}$.

Since the original channel takes twice as many channel uses to implement this scheme, the actual secrecy rate is half the value indicated by (43). However, this still means a positive secrecy rate is achievable.

This means that utilizing feedback signals leads to higher achievable secrecy rate for this channel.

### B. Outer Bound

Although we have shown that using feedback can improve the secrecy rate, it remains unclear whether this can only be done by letting Node $1$ use the signal $Y$ to compute $X$. If the signal $Y$ is not available to Node $1$, is it possible to achieve the same rate via a smarter way to compute $X_r$ at Node $2$? Additionally, if ignoring $Y$ at Node $1$ is suboptimal, is it possible to bound the consequent rate loss? To answer these questions, clearly, we need an outer bound on the secrecy capacity region of this model.

We begin by deriving a bound on $R_1$.

*Theorem 1:* For the channel model in Figure 1, $R_1$ is upper bounded by

$$\max_{\Pr(X,X_f)} \min\{I(X;Y), I(X;Y|Z,X_f) + I(X_f;Y_f,Z|X)\} \tag{44}$$

*Proof:* See Appendix A. ■

*Remark 3:* Ignoring $Y_f$ at Node $1$ is equivalent to viewing $Y_f$ as a constant. From (44), $R_1$, in this case, is upper bounded by

$$\max_{\Pr(X,X_f)} \min\{I(X;Y), I(X;Y|Z,X_f) + I(X_f;Z|X)\} \tag{45}$$

which is the upper bound proved in [24]. □

*Theorem 2:* The secrecy capacity region of the channel model in Figure 1 is bounded by

$$\cup_{\Pr\left(X,X_f\right)} \{(R_1,R_2) : \text{(47) (48) (49) holds}\} \tag{46}$$

$$0 \le R_1 \le I\left(X;Y\right) \tag{47}$$

$$0 \le R_2 \le I\left(X_f;Y_f\right) \tag{48}$$

$$R_1 + R_2 \le \min \left\{ \begin{array}{l} I\left(X;Y|Z,X_f\right) + I\left(X_f;Z,Y_f|X\right), \\ I\left(X_f;Y_f|Z,X\right) + I\left(X;Z,Y|X_f\right) \end{array} \right\} \tag{49}$$

*Proof:* The proof is provided in Appendix B. ■

For a deterministic binary wire-tap channel, Theorem 2 leads to the equivocation capacity region, as shown by the following theorem:

*Theorem 3:* When $X, X_f$ are binary and $Y = X \oplus X_f, Y_f = X_f \oplus X, Z = X \oplus X_f$, the secrecy capacity region is given by

$$R_j \geq 0, j = 1, 2 \tag{50}$$

$$R_1 + R_2 \leq 1 \tag{51}$$

*Proof:* The achievability follows from [26, Theorem 2]. The converse follows from Theorem 2. The sum rate bound specializes as follows:

$$I\left(X; Y|Z, X_f\right) + I\left(X_f; Y_f, Z|X\right) \tag{52}$$

$$= I\left(X; X|X \oplus X_f, X_f\right) + I\left(X_f; X_f, X \oplus X_f|X\right) \tag{53}$$

$$= I\left(X; X|X, X_f\right) + I\left(X_f; X_f, X|X\right) \tag{54}$$

$$= I\left(X_f; X_f, X|X\right) \tag{55}$$

$$\leq H\left(X_f\right) \leq 1 \tag{56}$$

■

We next consider the Gaussian channel.

*Theorem 4:* When $Y_f$ is a constant, i.e., $Y_f$ is ignored by Node $1$, the secrecy rate $R_1$ is upper bounded by

$$\inf_{\sigma^2 \geq 0} C\left(\frac{P\left(1+\sigma^2-\sqrt{h_1}\rho\right)^2}{\left(1+\sigma^2-\rho^2\right)\left(h_1 P+1+\sigma^2\right)}\right) + C\left(\frac{h_2 P_r}{1+\sigma^2}\right) \tag{57}$$

*Proof:* Define $N_4$ as a Gaussian random variable such that $N_4 \sim \mathcal{N}(0, \sigma^2)$ and is independent from $N_i, i = 1, 2, 3$. Recall that $Z$ is the signal received by the eavesdropper. We next consider a channel where the eavesdropper receives $Z + N_4$. Since $Z + N_4$ is a degraded version of $Z$, we can find an upper bound of the original channel by deriving an upper bound for this new channel. This upper bound is found by applying the bound (45).

We next prove that all terms in the upper bound (45) is maximized when $X, X_f$ are independent and each has a Gaussian distribution with zero mean and maximum possible variance: $I(X; Y)$ is obviously maximized by this distribution. For the other two terms, we have:

$$I\left(X; Y|X_f, Z\right) \tag{58}$$

$$=I\left(X;X+N_1|X_f,\sqrt{h_1}X+\sqrt{h_2}X_f+N_2+N_4\right) \quad (59)$$

$$=h\left(X+N_1|X_f,\sqrt{h_1}X+\sqrt{h_2}X_f+N_2+N_4\right)-h\left(N_1|N_2+N_4\right) \quad (60)$$

$$\leq h\left(X+N_1|\sqrt{h_1}X+N_2+N_4\right)-h\left(N_1|N_2+N_4\right) \quad (61)$$

and

$$I\left(X_f;Z|X\right) \quad (62)$$

$$=I\left(X_f;\sqrt{h_2}X_f+N_2+N_4|X\right) \quad (63)$$

$$=h\left(\sqrt{h_2}X_f+N_2+N_4|X\right)-h\left(N_2+N_4\right) \quad (64)$$

$$\leq h\left(\sqrt{h_2}X_f+N_2+N_4\right)-h\left(N_2+N_4\right) \quad (65)$$

Equations (61) and (65) show that the second term in (45) is maximized when $X$ and $X_f$ are independent. Moreover, (61) is known to be maximized when $X$ has a Gaussian distribution with the maximum possible variance; see [27]. (65) is also maximized when $X_f$ has a Gaussian distribution with the maximum possible variance. Hence we have shown the optimal input distribution for $X, X_f$ is an independent Gaussian distribution. For this distribution, it can be verified the second term in (45) becomes (57).

Hence we have proved the theorem. ■

*Remark 4:* When $\sigma^2 \to \infty$, (57) converges to $C(P)$, which corresponds to the first term in (45). Thus, (57) is written as one term instead of the two terms as in (45). □

*Remark 5:* We introduce $N_4$ to further tighten the bound. For example, consider the case where $\rho = \eta = 0$. In this case the upper bound can be expressed as

$$\min_{0\leq\alpha\leq1} C\left(\frac{P}{\alpha h_1 P+1}\right)+C\left(\alpha h_2 P_r\right) \quad (66)$$

where $\alpha = 1/(1+\sigma^2)$. Consider choosing the remaining parameters as $h_1 = 1, h_2 = 10$, $P = 100, P_r = 5$. It can be verified that the minimum is attained around $\alpha = 0.09$, and not at $\sigma^2 = 0$. Hence, the bound presented here is tighter than the bound in [24]. □

Next, we present the following theorem.

*Theorem 5:* The secrecy capacity region of the Gaussian two-way wiretap channel is outer

bounded by

$$0 \leq R_1 \leq C(P) \tag{67}$$

$$0 \leq R_2 \leq C(P_r) \tag{68}$$

$$R_1 + R_2 \leq \min \left\{ \begin{array}{l} \inf_{\sigma^2 \geq 0} C\left(\frac{P\left(1+\sigma^2-\sqrt{h_1}\rho\right)^2}{(1+\sigma^2-\rho^2)(h_1P+1+\sigma^2)}\right) + C\left(\frac{P_r\left(h_2+1+\sigma^2-2\sqrt{h_2}\eta\right)}{1+\sigma^2-\eta^2}\right) \\ \inf_{\sigma^2 \geq 0} C\left(\frac{P_r\left(1+\sigma^2-\sqrt{h_2}\eta\right)^2}{(1+\sigma^2-\eta^2)(h_2P_r+1+\sigma^2)}\right) + C\left(\frac{P\left(h_1+1+\sigma^2-2\sqrt{h_1}\rho\right)}{1+\sigma^2-\rho^2}\right) \end{array} \right\} \tag{69}$$

*Proof:* Again we consider a channel where the eavesdropper receives $Z + N_4$ and derive an outer bound for this new channel. $N_4$ is as defined in the proof of Theorem 4.

To prove the theorem, we first show $I(X;Y)$, $I(X;Y|Z,X_f)$, $I(X_f;Y_f,Z|X)$, $I(X_f;Y_f)$, $I(X_f;Y_f|Z,X_f)$ and $I(X;Z,Y|X_f)$ are maximized simultaneously when $X$ and $X_f$ are independent, $X \sim \mathcal{N}(0,P)$, and $X_f \sim \mathcal{N}(0,P_r)$.

Due to the symmetry of the channel model, we only need to show $I(X;Y)$, $I(X;Y|Z,X_f)$ and $I(X_f;Y_f,Z|X)$ are maximized by this distribution.

The case of $I(X;Y|Z,X_f)$ was shown in the proof of Theorem 4.

For $I(X_f;Y_f,Z|X)$, we have:

$$I\left(X_f;Y_f,Z|X\right) \tag{70}$$

$$=I\left(X_f;X_f+N_3,\sqrt{h_2}X_f+N_2+N_4|X\right) \tag{71}$$

$$=h\left(\sqrt{h_2}X_f+N_2+N_4,X_f+N_3|X\right)-h\left(N_2+N_4,N_3\right) \tag{72}$$

$$\leq h\left(\sqrt{h_2}X_f+N_2+N_4,X_f+N_3\right)-h\left(N_2+N_4,N_3\right) \tag{73}$$

Hence $I(X_f;Y_f,Z|X)$ is maximized when $X$ and $X_f$ are independent, $X \sim \mathcal{N}(0,P)$, and $X_f \sim \mathcal{N}(0,P_r)$. The theorem then is a consequence of Theorem 2 when evaluated at this input distribution. ■

*Remark 6:* The introduction of $N_4$ is again useful in tightening the bound. For example, consider the case where $\rho = \eta = 0$, $h_1 = 1, h_2 = 10$, $P = 100, P_r = 5$.

In this case the upper bound on $R_1$, which is $C(P)$, is about $3.3291$. The first term inside the minimum in (69), which is also an upper bound on $R_1$ takes the form:

$$\min_{0\leq\alpha\leq 1} C\left(\frac{P}{\alpha h_1 P+1}\right)+C\left(P_r\left(\alpha h_2+1\right)\right) \tag{74}$$

where $\alpha = 1/(1+\sigma^2)$. It can be verified that the minimum is smaller than $3.24$ and is attained around $\alpha = 0.32$. Hence the upper bound on $R_1$ is dominated by the first term inside the minimum in (69) and is not attained at $\sigma^2 = 0$. □

### *C. Achievable Rates for the Gaussian Two-way Wiretap Channel*

Let us use $[x]^+$ to denote $\max\{x, 0\}$. Then we have the following theorem.

*Theorem 6:* Define $R_1^*$ as

$$R_1^* = \max_{0\leq\alpha\leq 1} \alpha \left[C\left(P\right) - \left[C\left(\frac{h_1 P}{h_2 P_r + 1}\right) - \frac{1-\alpha}{\alpha}\left[C\left(P_r\right) - C\left(\frac{h_2 P_r}{h_1 P + 1}\right)\right]^+\right]^+\right]^+ \tag{75}$$

and $R_2^*$ as

$$R_2^* = \max_{0\leq\alpha\leq 1} \alpha \left[C\left(P_r\right) - \left[C\left(\frac{h_2 P_r}{h_1 P + 1}\right) - \frac{1-\alpha}{\alpha}\left[C\left(P\right) - C\left(\frac{h_1 P}{h_2 P_r + 1}\right)\right]^+\right]^+\right]^+ \tag{76}$$

Define the region $\mathbf{R}$ as the convex hull of the following three rate pairs of $(R_1, R_2)$:

$$(0,0), \quad (R_1^*, 0), \quad (0, R_2^*) \tag{77}$$

The rate region $\mathbf{R}$ is achievable.

*Proof:* The proof is given in Appendix C. ■

*Remark 7:* The achievable scheme is composed of two phases. During phase one, with a time sharing factor of $1-\alpha$, Node $2$ sends a key to Node $1$. During phase two, Node $1$ utilizes this key to encrypt its message and transmits the result to Node $2$. Hence when $\alpha = 1$, $\mathbf{R}$ is achieved when both nodes ignore their received signals when computing their transmitting signals. □

*Remark 8:* The achievable secrecy rate derived here may be potentially improved further by combining it with the scheme in Section III-A. However, as we shall see later, Theorem 6 is sufficient to bound the rate loss when the feedback signals are not used by the legitimate nodes. □

### *D. Comparing the achievable rates and the outer bound*

We first consider the case with independent link noise, which is the model considered in [12].

*1) $\rho = \eta = 0$:*

*Theorem 7:* When $\rho = \eta = 0$, $P_r = kP$, $k$ is a positive constant, and $h_j \neq 0, j = 1, 2$, the loss in secrecy rates when received signals are not used to compute transmitting signals at Node $j$, $j = 1, 2$ is bounded by a constant, which is only a function of $h_1$ and $h_2$.

*Proof:* The proof is given in Appendix D. ■

*Theorem 8:* Even in the case where cooperative jamming is possible ($h_j \neq 0, j = 1, 2$), when $P$ is not proportionally increasing with $P_r$, ignoring $Y_f$ at Node 1 can lead to unbounded loss in the secrecy rate.

*Proof:* The proof is given in Appendix E. ■

We next consider a special case of the model that attracted some interest in the past [24], [28]. In this model, $Z$ is a degraded version of $Y$ given $X_f$, and $Y_f$ is ignored by Node 1:

*2) $h_1 \leq 1, \rho = \sqrt{h_1}$ and $Y_f$ is a constant:* In this case, $N_2$ can be written as $\sqrt{h_1}N_1 + N_2'$, where $N_2'$ is independent from $N_1, N_3$ and $N_2' \sim \mathcal{N}(0, 1 - h_1)$. Then the signals received by the eavesdropper $Z$ can be expressed as:

$$Z = \sqrt{h_1}X + \sqrt{h_2}X_f + \sqrt{h_1}N_1 + N_2' \tag{78}$$

$$= \sqrt{h_1}\,(X + N_1) + N_2' + \sqrt{h_2}X_f \tag{79}$$

From this, we observe that, given $X_f$, $Z$ is a degraded version of $Y = X + N_1$.

*Corollary 1:* When $h_1 \leq 1, \rho = \sqrt{h_1}$, and $h_2 \neq 0$, $Y_f$ is a constant, then the achievable rate of $R_1$ using cooperative jamming is at most $0.5$ bit per channel use from the secrecy capacity.

*Remark 9:* Corollary 1 was first proposed in [28] and later appeared in [24]. Here we first describe the approach of [24]:

From Theorem 6 and Remark 7, the achievable rate for $R_1$ in this case is obtained by letting $\alpha = 1$ and evaluating $R_1^*$. In this case

$$R_1 = C(P) - C(\frac{h_1 P}{h_2 P_r + 1}) \tag{80}$$

The upper bound proposed in [24] on $R_1$ is

$$\min\{C(P), C(P) - C(h_1 P) + C(h_2 P_r)\} \tag{81}$$

Here we observe (81) can be obtained from (57) when evaluated with $\sigma^2 \to \infty$ and $\sigma^2 = 0$. Reference [24] proves Corollary 1 by comparing (80) and (81). It can be then verified that the gap between (80) and (81) is less than $0.5$ bit per channel use.

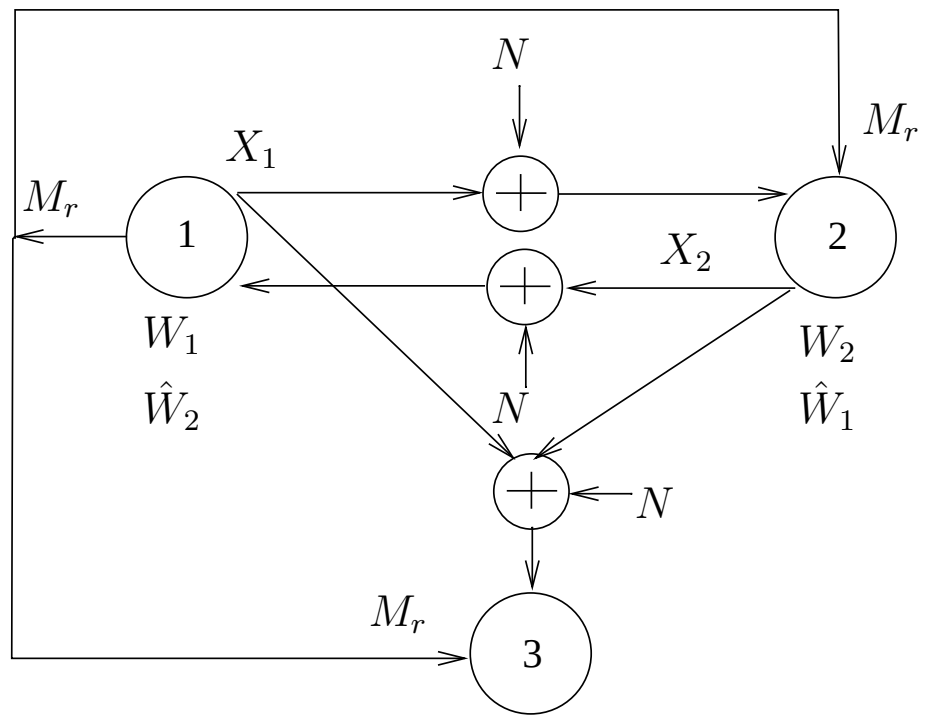

Fig. 7. Two-way wiretap Channel with Additional Public Noiseless Forward Link

The approach of reference [28] is different and uses results on a wiretap channel with noisy feedback. This proof is delegated to Section V.

## IV. Feedback in Half-Duplex Two-way Relay Channel with An Untrusted Relay

In this section, we derive the outer bound for the secrecy capacity region of the two-way relay channel with an untrusted relay in Section II-B (Figure 3). To find the outer bound, we first consider the channel in Figure 7.

We assume $X_1$ and $X_2$ have the same power constraint as the $X_1, X_2$ in Figure 3. $M_r$ is now accessible to Node 1 and delivered to the other nodes via a *public noiseless link*. The remaining part of the channel is activated when the original two-way relay channel is in the MAC mode, and is inactive when the original two-way relay channel model is in the broadcast mode. Doing so ensures the overall number of channel uses to be the same between these two models.

Recall that $M_j, j = 1, 2$ still models the local randomness at Node $j, j = 1, 2$. The encoding function of Node 1 at the $i$th channel use when the channel is active can be defined as:

$$X_{1,i} = \tilde{f}_{1,i}(Y_1^{i-1}, W_1, M_1, M_r) \tag{82}$$

Similarly, the encoding function of Node 2 at the $i$th channel use when the channel is active can be defined as:

$$X_{2,i} = \tilde{f}_{2,i}(Y_2^{i-1}, W_2, M_2, M_r) \tag{83}$$

With these preparations, we present the following theorem:

*Theorem 9:* The secrecy rate region of the channel in Figure 7 includes the secrecy capacity region of the two-way relay channel in Figure 3.

*Proof:* Consider the model in Figure 3. Suppose during a MAC mode, a genie reveals $X_1 + X_2 + N$ to Node 1 and Node 2. We also add a public noiseless link that takes inputs from Node 1 and provides outputs to Node 2 and the relay. We make $M_r$ accessible to Node 1 and use the public noiseless link to deliver $M_r$ to Node 2 and the relay. This side information does not increase the knowledge of the relay and hence will not decrease the secrecy capacity region of the channel.

During a broadcast mode, a genie reveals the link noise level $N_2$ to Node 2. Similarly, the link noise $N_1$ is revealed to Node 1. This side information will not decrease the secrecy capacity region of the channel either.

With the side information provided to the nodes, the links from the relay to Node $1, 2$ can be removed. This is because

1) Node 1 and Node 2 have the signal received by the relay $X_1 + X_2 + N$.
2) Node 1 sends $M_r$ via the public noiseless forward link. With $M_r$ available at Node 2, it can compute the signal transmitted by the relay node. Due to the same reason, Node 1 knows the signal transmitted by the relay node as well.
3) With noise $N_2$ available at Node 2, Node 2 can compute the signal it received from the relay. For similar reasons, Node 1 can compute the signal it received from the relay as well.

Since $N_1, N_2, N$ are independent, $N_1$ and $N_2$ can be incorporated as the local randomness at Node 1 and Node 2 respectively.

After removing the links from the relay to Node $1, 2$, the channel indeed becomes that which is described by Figure 7, where Node 3 corresponds to the relay node whose output broadcast link to Node $1, 2$ is removed. Since, every step we took during this transformation could only expand the secrecy capacity region, we have proved the theorem. ■

To derive an outer bound for the secrecy capacity of the channel in Figure 7, we first consider the case when the channel is active regardless of whether the two-way relay channel is in MAC mode or broadcast mode. We recognize that in this case, the channel becomes a special case of the two-way wiretap channel defined in Section II. Utilizing this connection leads to the following corollary:

*Corollary 2:* The secrecy capacity region of the channel in Figure 7 is outer bounded by

$$R_1 + R_2 \leq \min\left\{C\left(\bar{P}_1\right), C\left(\bar{P}_2\right)\right\} \tag{84}$$

$$R_1 \geq 0, R_2 \geq 0 \tag{85}$$

where $\bar{P}_i$ is the average power constraint of Node $i$.

*Proof:* The channel in Figure 7 is a special case of the channel defined in (1), where

$$Y, Y_f, Z, X, X_f \tag{86}$$

correspond to

$$\{X_1 + N, M_r\}, X_2 + N, \{X_1 + X_2 + N, M_r\}, \{X_1, M_r\}, X_2 \tag{87}$$

respectively, and $\Pr(Y, Y_f, Z|X, X_f)$ becomes $\Pr(N)$.

Therefore the corollary follows as a direct consequence of Theorem 5 with $\eta = \rho = 1$, $h_1 = h_2 = 1$, $\sigma^2 = 0$. ■

Note that to apply Corollary 2 to the half-duplex two-way relay channel, we need to take into account the channel uses when the channel in Figure 7 is inactive during the channel uses when the original two-way relay channel is in the broadcast mode. Hence, the outer bound in Corollary 2 becomes the following region $\mathbf{A}$:

$$R_1 + R_2 \leq \alpha \min\left\{C\left(\bar{P}_1/\alpha\right), C\left(\bar{P}_2/\alpha\right)\right\} \tag{88}$$

$$R_1 \geq 0, R_2 \geq 0 \tag{89}$$

which reflects the number of channel uses during which some nodes are inactive.

Define region $\mathbf{B}$ as

$$0 \leq R_1 \leq (1-\alpha)C(\bar{P}_r/(1-\alpha)) \tag{90}$$

$$0 \leq R_2 \leq (1-\alpha)C(h\bar{P}_r/(1-\alpha)) \tag{91}$$

Then we have the following theorem:

*Theorem 10:* An outer bound for the secrecy capacity of two-way relay channel is given by

$$\cup_{0 \leq \alpha \leq 1}\{\mathbf{A} \cap \mathbf{B}\} \tag{92}$$

*Proof:* Region $\mathbf{A}$ follows by applying Corollary 2 and taking into account the fact that the channel is inactive when the original two-way relay channel is in broadcast mode as described above.

Region $\mathbf{B}$ follows from removing the secrecy constraint and applying the cut-set bound in [10, Theorem 15.10.1]. To derive (91), we consider the cut where the set $T$ includes the relay node and Node 2. From the cut-set bound, we get:

$$H(W_2) \leq mI(X_2, X_r; Y_1|X_1) + (m+n)\varepsilon \tag{93}$$

$\varepsilon > 0$ and $\lim_{m+n\to\infty} \varepsilon = 0$.

Therefore, we get

$$\frac{1}{m+n}H(W_2) \leq \frac{m}{m+n}I(X_2, X_r; Y_1|X_1) + \varepsilon \tag{94}$$

It is easy to see that for the Gaussian two-way relay channel, $I(X_2, X_r; Y_1|X_1)$ is maximized when $X_1, X_2, X_r$ takes an independent Gaussian distribution with maximum possible variance. Let $m+n \to \infty$, and use the fact that $\lim_{m+n\to\infty} \frac{m}{m+n} = 1-\alpha$, we obtain (91) by evaluating (94) for this distribution.

Equation (90) is derived similarly due to the symmetry of the channel model.

Hence we proved the theorem. ■

*Remark 10:* When $\bar{P}_r \to \infty$, and $h \neq 0$, then the region is maximized when $\alpha \to 1$. The outer bound becomes:

$$R_1 + R_2 \leq \min\left\{C\left(\bar{P}_1\right), C\left(\bar{P}_2\right)\right\} \tag{95}$$

$$R_1 \geq 0, R_2 \geq 0 \tag{96}$$

□

### A. Comparison with Achievable Rates

In this section, we compare the outer bound with the achievable secrecy rate region.

We begin by restating an achievable rate for $R_1$ from [17]. The rate region then follows from time sharing.

*Theorem 11:* [17, Theorem 1] The following secrecy rate of $R_1$ is achievable for the model in Figure 3:

$$0 \leq R_1 \leq \max_{0\leq P_1'\leq \bar{P}_1/\alpha, 0<\alpha<1} \alpha\left[C\left(\frac{P_1'}{(1+\sigma_c^2)}\right) - C\left(\frac{P_1'}{(1+P_2)}\right)\right]^+ \tag{97}$$

where $\sigma_c^2$ is the variance of the Gaussian quantization noise determined by:

$$\alpha C\left(\frac{P_1'+1}{\sigma_c^2}\right) = (1-\alpha)C\left(P_r\right) \tag{98}$$

$P_2$ was defined in (25), $P_r$ was defined in (26).

*Remark 11:* The achievable scheme above uses compress-and-forward. And Node $1$ and $2$ ignore their received signals when computing the transmitted signals. The proof can be found in [17].

*Remark 12:* For any fixed $\alpha$ such that $0 < \alpha < 1$, if the power of the relay $\bar{P}_r \to \infty$, then $\sigma_c^2 \to 0$, the achievable rate converges to

$$\alpha(C(P_1) - C(\frac{P_1}{1+P_2})) \tag{99}$$

Equation (99) is a monotonic increasing function of $\alpha$. Hence, as long as $\alpha < 1$, we can always increase $\alpha$ and increase the achievable secrecy rate. Therefore, when $\bar{P}_r \to \infty$, the optimal time sharing factor $\alpha \to 1$. The achievable rate then converges to

$$C(\bar{P}_1) - C(\frac{\bar{P}_1}{1+\bar{P}_2}) \tag{100}$$

The secrecy rate region is obtained with time sharing and it converges to

$$R_1 + R_2 \le C(\bar{P}_1) - C(\frac{\bar{P}_1}{1+\bar{P}_2}) \tag{101}$$

$$R_1 \ge 0, R_2 \ge 0 \tag{102}$$

□

Utilizing this result, we have the following corollary:

*Corollary 3:* When $\bar{P}_r \to \infty$, the gap between the outer bound and the achievable rate is bounded by $0.5$ bit per channel use.

To prove this corollary, we need the following lemma:

*Lemma 1:* Define the following functions:

$$f(x,y) = \frac{1}{2}\log_2\left(\frac{(1+x)(1+y)}{1+x+y}\right) \tag{103}$$

$$g(x,y) = \min\{C(x), C(y)\} \tag{104}$$

Let $h(x,y) = g(x,y) - f(x,y)$. Then $0 \le h(x,y) \le 0.5$.

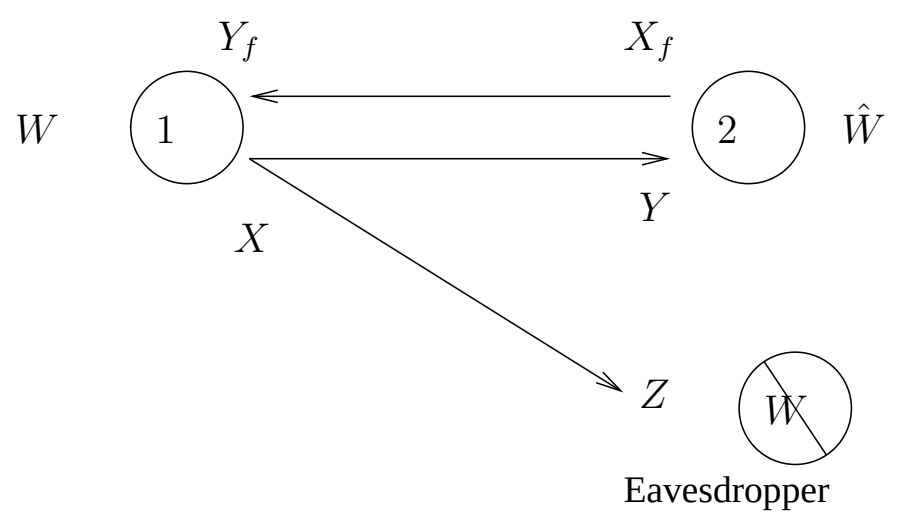

Fig. 8. A wiretap channel with noisy feedback

*Proof:* Without loss of generality $x \le y$. For $x \ge y$, simply exchange $x$ and $y$. $h(x,y)$ is given by

$$h(x,y) = \frac{1}{2}\log_2\left(\frac{1+x+y}{1+y}\right) \tag{105}$$

$$= \frac{1}{2}\log_2\left(1+\frac{1+x}{1+y}\right) \tag{106}$$

$$\le \frac{1}{2}\log_2(1+1) = 0.5 \tag{107}$$

Clearly $h(x,y) \ge 0$. Hence $0 \le h(x,y) \le 0.5$. ■

Corollary 3 can then be proved by letting $x = \bar{P}_1, y = \bar{P}_2$. The upper bound on the sum rate and the achievable sum secrecy rate then become $g(x,y)$ and $f(x,y)$ when $\bar{P}_r \to \infty$. Using Lemma 1 we prove the gap between the upper bound and lower bound of the sum secrecy rate is less than 0.5 bit per channel use. Since the achievable region and the outer bound are only different on the bounds for the sum rate, this proves the gap between the inner bound and outer bound of the secrecy capacity region is also less than 0.5 bit per channel use when $\bar{P}_r \to \infty$. Hence we have proved Corollary 3.

## V. Alternative Proofs of Corollary 1 and Corollary 2

Corollary 1 and Corollary 2 can also be proved by using results on the wiretap channel with noisy feedback [28]. In this section we provide these proofs for completeness.

### A. A Wiretap channel with noisy feedback

The channel model is shown in Figure 8. Node $1$ sends a message $W$ to Node $2$, which must be kept secret from the eavesdropper. The channel is described by

$$\Pr(Y,Z,Y_f|X,X_f) = \Pr(Y,Z|X)\Pr(Y_f|X_f) \tag{108}$$

Within each channel use, Node $1$ and Node $2$ *take turns to transmit*. This implies that, Figure 8 is *not* a special case of the two-way wiretap channel in Figure 1. Without loss of generality, we assume Node $2$ transmits first.

At the $i$th channel use, the encoding function of Node $1$ is defined as:

$$X_i = f_i(Y_f^i, W, M_1) \tag{109}$$

Note that since Node $2$ transmits first, Node $1$ has an extra sample of $Y_f$ to use when computing its transmitted signals. Therefore in (109), $Y_f^i$ is used instead of $Y_f^{i-1}$.

The encoding function of Node $2$ is defined as

$$X_{f,i} = g_i(Y^{i-1}, M_2) \tag{110}$$

$f_i, g_i$ are deterministic functions.

Let $n$ be the total number of channel uses. The destination must decode $W$ reliably from $X_f^n, Y^n, M_2$. Hence from Fano's inequality, we have

$$H(W|X_f^n, Y^n, M_2) < n\varepsilon_4 \tag{111}$$

where $\varepsilon_4 > 0$ and $\lim_{n\to\infty} \varepsilon_4 = 0$.

The message $W$ must be kept secret from the eavesdropper. Hence

$$I(W; Z^n) < n\varepsilon_5 \tag{112}$$

where $\varepsilon_5 > 0$ and $\lim_{n\to\infty} \varepsilon_5 = 0$.

*Theorem 12:* The secrecy capacity of the channel model in Figure 8 is upper bounded by

$$R_e \leq \max_{\Pr(X,X_f)} \min\{I(X;Y), I(X;Y|Z) + I(X_f;Y_f)\} \tag{113}$$

*Proof:* The proof is provided in Appendix F. ■

*Corollary 4:* If $X - Y - Z$ is a Markov chain, the secrecy capacity of the channel model in in Figure 8 is given by

$$\max_{\Pr(X,X_f)} \min\{I(X;Y), I(X;Y) - I(X;Z) + I(X_f;Y_f)\} \tag{114}$$

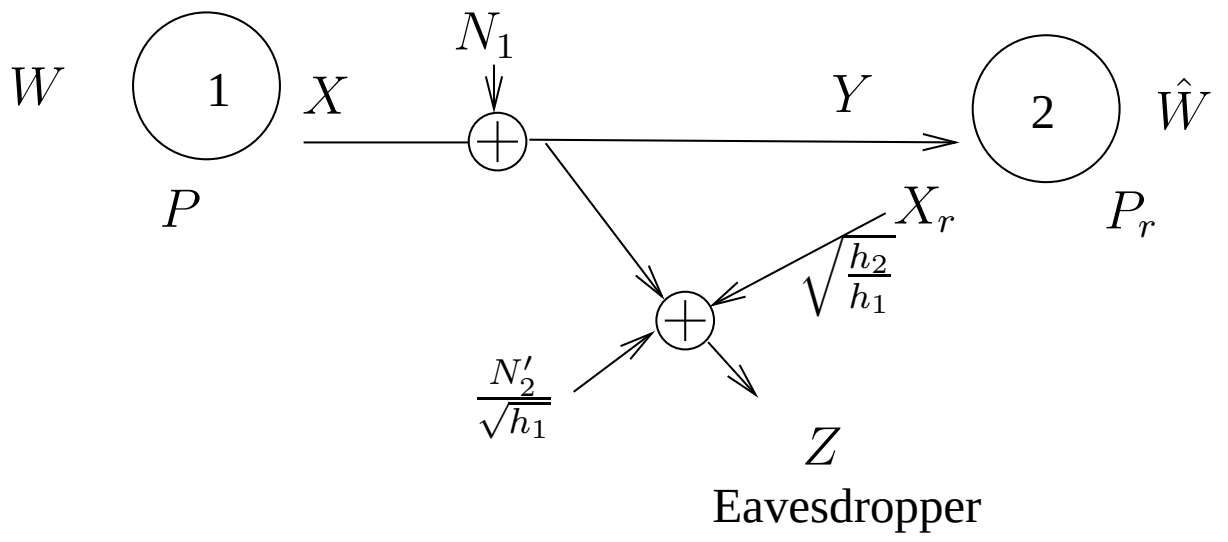

Fig. 9. The degraded model

*Proof:* Equation (114) is achievable because of [8, Theorem 3.1]. In details, the rate (114) can be obtained by letting $V_f = X$, $Y_f = Y$, $Z_f = Z$, $U_f = \phi$, $V_b = X_f$, $Y_b = Y_f$, $Z_b = \phi$ in [8, Theorem 3.1].

When $X - Y - Z$ is a Markov chain, we notice $I(X;Y|Z) = I(X;Y) - I(X;Z)$. Hence the achievable (114) matches the upper bound in (113). ■

*Remark 13:* When the backward channel $\Pr(Y_f|X_f)$ is a rate limited noiseless link, whose rate is $R_f$, then $I(X_f;Y_f) = H(X_f) = R_f$. Then using Corollary 4, we obtain the result in [5]. □

### B. Alternative Proof of Corollary 1

*Theorem 13:* For the degraded case considered in Section III-D2, $R_1$ is upper bounded by

$$\min\{C(P), C(\frac{h_2}{h_1}\bar{P}_r)\} \tag{115}$$

where $\bar{P}_r = P_r + \frac{1-h_1}{h_2}$.

*Remark 14:* Since $h_1 \leq 1$, it can be verified that the upper bound in (115) is looser than (81). However, this bound is sufficient to prove the $0.5$ bit gap result.

*Proof:*

We begin by redrawing the channel model in Figure 9, where $N_2'$ is a zero mean Gaussian random variable with variance $1 - h_1$.

The first term $C(P)$ follows by removing the eavesdropper and applying the upper bound for Gaussian two-way channel from [2].

In order to obtain the second term in (115), we convert the model in Figure 9 to the model in Figure 10. In this new model, $N_2'$ is removed, and the power constraint of Node $2$ is increased from $P_r$ to $\bar{P}_r = P_r + \frac{1-h_1}{h_2}$. Since Node $2$ can always use this additional power to transmit noise which is statistically equivalent to $N_2$, the secrecy capacity of Figure 10 is greater or equal to the secrecy capacity of Figure 9.

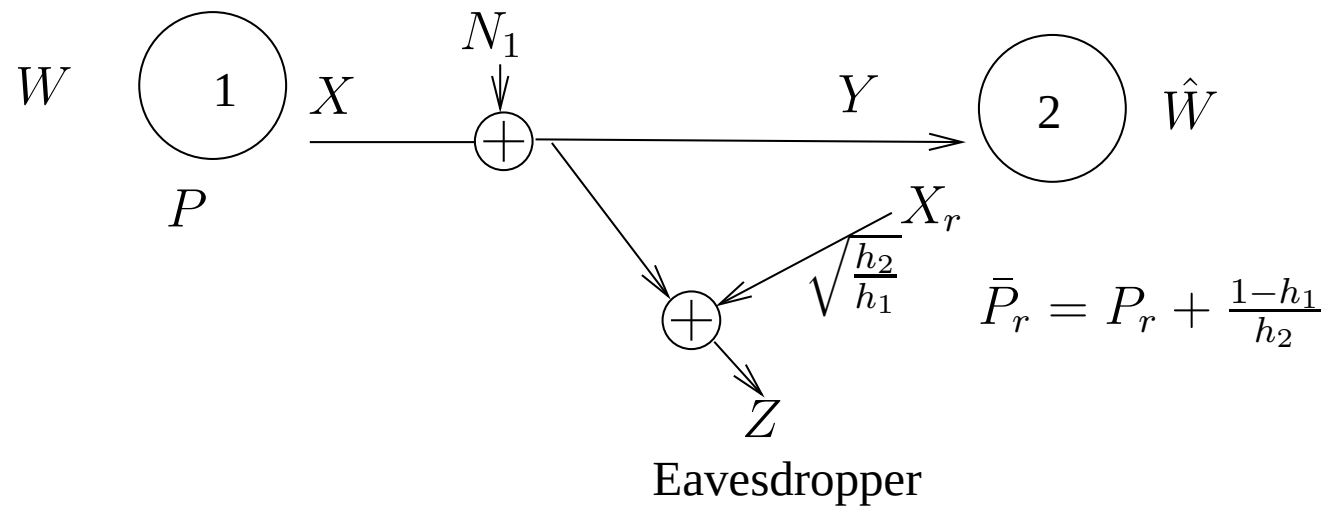

Fig. 10. The model with cooperative jamming

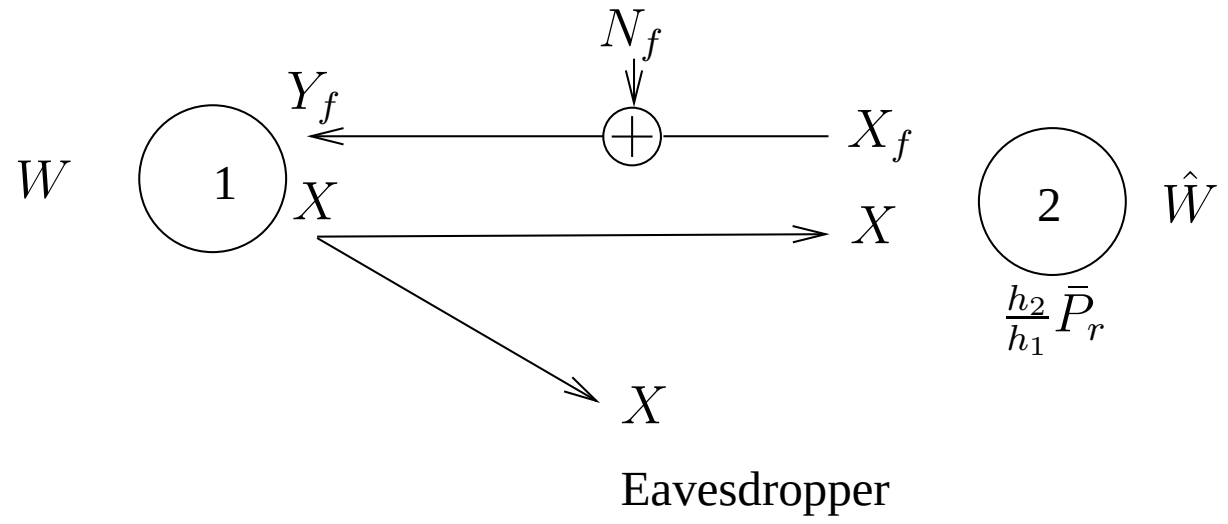

Fig. 11. The model with feedback

We next prove that the secrecy capacity of the model in Figure 11 must be greater or equal to the secrecy capacity of the channel model in Figure 10. Figure 11 is a special case of the wiretap channel with noisy feedback. For this model, the encoding functions used by Node $1$ and $2$ at the $i$th channel use are denoted by $f_i$ and $g_i$ and are defined in (109) and (110). The power constraint of Node $2$ is $\frac{h_2}{h_1}\bar{P}_r$. Node $1$ is not constrained in transmission power.

We prove that any signaling scheme in Figure 10 can be simulated by Figure 11. This means for any set of encoding functions of Node $1$ and $2$, denoted by $\{\tilde{f}_i\}$ $\{\tilde{g}_i\}$ respectively in Figure 10, we can find encoding functions $\{f_i\}$, $\{g_i\}$ for Figure 11, such that given the same noise sequence $N_f = N_1$ and the same local randomness $M_j, j = 1, 2$, the message can be reliably received by Node $2$ at the same secrecy rate. This can be shown as follows:

We choose $f_i$, the encoding function used by Node $1$ in Figure 11 as:

$$X_i = Y_{f,i} + \tilde{f}_i\left(W, M_1\right) \tag{116}$$

$g_i$, the encoding function used by Node $2$ in Figure 11 is chosen as:

$$X_{f,i} = \sqrt{\frac{h_2}{h_1}} \tilde{g}_i(X^{i-1} - X_f^{i-1}, M_2) \tag{117}$$

Then, as shown below, if $X_f^{i-1} = \sqrt{\frac{h_2}{h_1}} X_r^{i-1}$, then $X_{f,i} = \sqrt{\frac{h_2}{h_1}} X_{r,i}$. The notation $\tilde{f}^{i-1}(M_1, W)$ stands for $\tilde{f}_j(M_1, W), j = 1, ..., i-1$.

We begin with:

$$X_{f,i} = \sqrt{\frac{h_2}{h_1}} \tilde{g}_i \left( X^{i-1} - X_f^{i-1}, M_2 \right) \tag{118}$$

Using (116), we get:

$$X_{f,i} = \sqrt{\frac{h_2}{h_1}} \tilde{g}_i \left( Y_f^{i-1} + \tilde{f}^{i-1} (W, M_1) - X_f^{i-1}, M_2 \right) \tag{119}$$

$$= \sqrt{\frac{h_2}{h_1}} \tilde{g}_i \left( X_f^{i-1} + N_f^{i-1} + \tilde{f}^{i-1} (W, M_1) - X_f^{i-1}, M_2 \right) \tag{120}$$

$$= \sqrt{\frac{h_2}{h_1}} \tilde{g}_i \left( N_f^{i-1} + \tilde{f}^{i-1} (W, M_1), M_2 \right) \tag{121}$$

Since $N_f^{i-1} = N_1^{i-1}$, (121) equals:

$$\sqrt{\frac{h_2}{h_1}} \tilde{g}_i \left( N_1^{i-1} + \tilde{f}^{i-1} (W, M_1), M_2 \right) \tag{122}$$

Since $X^{i-1} = \tilde{f}^{i-1}(W, M_1)$, (122) equals

$$\sqrt{\frac{h_2}{h_1}} \tilde{g}_i \left( N_1^{i-1} + X^{i-1}, M_2 \right) \tag{123}$$

$$= \sqrt{\frac{h_2}{h_1}} \tilde{g}_i \left( Y^{i-1}, M_2 \right) \tag{124}$$

$$= \sqrt{\frac{h_2}{h_1}} X_{r,i} \tag{125}$$

Hence, when $n$ channel uses are involved, we have $X_f^n = \sqrt{\frac{h_2}{h_1}} X_r^n$.

Using this result, from (116) we have

$$X_i = Y_{f,i} + \tilde{f}_i (W, M_1) \tag{126}$$

$$= X_{f,i} + N_{f,i} + \tilde{f}_i (W, M_1) \tag{127}$$

$$= \sqrt{\frac{h_2}{h_1}} X_{r,i} + N_{1,i} + \tilde{f}_i (W, M_1) \tag{128}$$

$$= \sqrt{\frac{h_2}{h_1}} X_{r,i} + N_{1,i} + X_i = Z_i \tag{129}$$

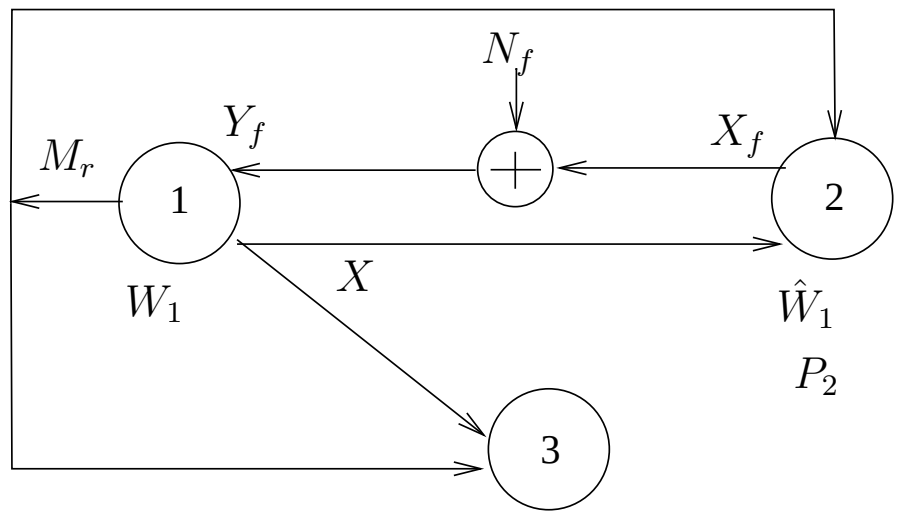

Fig. 12. Two-way model with one-sided secure link

Therefore the signals received by the eavesdropper in Figure 11 is the same signals received by the eavesdropper in Figure 10.

The destination in Figure 11 knows $X_{f,i}$. Therefore, it can compute $\tilde{f}_i(W, M_1) + N_{f,i}$ from $X_i - X_{f,i}$. On the other hand, $\tilde{f}_i(W, M_1) + N_{f,i} = \tilde{f}_i(W, M_1) + N_{1,i}$ is exactly the signal received by Node $2$ in Figure 10 at the $i$th channel use. This fact, along with the fact that $X_{r,i} = X_{f,i}$, shows that the destination in Figure 11 can compute any signal known by the destination in Figure 10. This means that, if $W$ can be reliably received in Figure 10, it can also be reliably received in Figure 11 at the same rate.

Hence we have proved that an upper bound for the secrecy capacity of Figure 11 is an upper bound for Figure 10. From Corollary 4 it follows that the secrecy capacity of Figure 11 is $C(\frac{h_2}{h_1}\bar{P}_r)$. Applying it to Figure 10, we obtained the second term in the upper bound (115). Hence we have the theorem. ■

The $0.5$ bit gap then follows from Lemma 1. In our case, the achievable rate can be expressed as $f(x, y)$ where $x = P$, $y = \frac{h_2}{h_1}\bar{P}_r$. The upper bound can be expressed as $h(x, y)$. Hence from Lemma 1, we proved the gap between the achievable rate (80) and the upper bound (115) can not exceed $0.5$ bits per channel use.

### *C. Alternative Proof of Corollary 2*

Consider the channel in Figure 12. It is the same channel as Figure 11 except that the power constraint of Node $2$ is changed to $P_2$. Again it is a a special case of the wiretap channel with noisy feedback. For this model, the encoding functions used by Node $1$ and $2$ at the $i$th channel use are denoted by $f_i$ and $g_i$ and are defined in (109) and (110). $W$ is replaced by $W_1$.

*Theorem 14:* The secrecy rate for the channel in Figure 12, where each node takes turn to transmit and Node $2$ transmits first, is greater than or equal to the maximal achievable individual rate $R_1$ of the channel in Figure 7.

*Proof:* We prove the theorem by showing any coding scheme of Figure 7 can be simulated by the channel in Figure 12. This means for any set of encoding functions of node $j$, $\{\tilde{f}_{j,i}\}$, $j = 1, 2$, in Figure 7, we can find encoding functions for Node $1$ and Node $2$ in Figure 12, such that at the same secrecy rate the message $W_1$ can be reliably received by Node $2$. For the encoding functions defined in (82) and (83), we choose the encoding functions for Figure 12 are chosen as:

$$X_i = \tilde{f}_{1,i}\left(Y_f^{i-1}, W_1, M_1, M_r\right) + Y_{f,i} \tag{130}$$

$$X_{f,i} = \tilde{f}_{2,i}\left(X^{i-1} - X_f^{i-1}, W_2, M_2, M_r\right) \tag{131}$$

$M_r$ is obtained by letting Node $1$ transmitting it over the noiseless public forward link. Then, with these encoding functions, the eavesdropper receives exactly the same signal as the signal received by the eavesdropper in Figure 7, if these two models experience the same noise sequence $N_f = N$ and the same local randomness $M_j, j = 1, 2, M_r$. This can be proved as follows:

We begin by assuming $X_f^{i-1} = X_2^{i-1}$, and prove $X_{f,i} = X_{2,i}$ and $X_i - Y_{f,i} = X_{1,i}$.

To prove $X_{f,i} = X_{2,i}$, we compare (131) with (83) and find that we need to prove $X^{i-1} - X_f^{i-1} = Y_2^{i-1}$. We begin with

$$X^{i-1} - X_f^{i-1} = X^{i-1} - Y_f^{i-1} + N_f^{i-1} \tag{132}$$

From (130), we have

$$X^{i-1} - Y_f^{i-1} \tag{133}$$

$$= \tilde{f}_1^{i-1}\left(Y_f^{i-1}, W_1, M_1, M_r\right) \tag{134}$$

Since we assume $X_f^{i-1} = X_2^{i-1}$ and $N_f^{i-1} = N^{i-1}$, we get

$$Y_f^{i-1} = X_f^{i-1} + N_f^{i-1} = X_2^{i-1} + N^{i-1} = Y_1^{i-1} \tag{135}$$

Hence (134) equals

$$\tilde{f}_1^{i-1}\left(Y_1^{i-1}, W_1, M_1, M_r\right) = X_1^{i-1} \tag{136}$$

The equality in (136) follows from (82). Hence $X^{i-1} - X_f^{i-1} = X_1^{i-1}$ and (132) equals:

$$X^{i-1} - X_f^{i-1} = X_1^{i-1} + N_f^{i-1} = X_1^{i-1} + N^{i-1} = Y_2^{i-1} \tag{137}$$

Hence from (137) we have shown

$$X_{f,i} = X_{2,i} \tag{138}$$

From (135), by comparing (130) with (82), we get

$$X_i - Y_{f,i} = X_{1,i} \tag{139}$$

From (138) and (139), we get

$$X_i = X_{1,i} + Y_{f,i} = X_{1,i} + X_{f,i} + N_{f,i} = X_{1,i} + X_{2,i} + N_i \tag{140}$$

Until this point, we have shown that the signals received by the eavesdroppers in the two models in Figure 12 and Figure 7 are identical.

From (139), we get

$$Y_{2,i} = X_{1,i} + N_i = X_i - Y_{f,i} + N_i = X_i - Y_{f,i} + N_{f,i} = X_i - X_{f,i} \tag{141}$$

Hence Node $2$ in Figure 12 can recover the signals received by Node $2$ in Figure 7. On the other hand, since $X_{f,i} = X_{2,i}$, Node $2$ in Figure 12 can also recover the signals transmitted by Node $2$ in Figure 7. This means if a message can be reliably decoded at a certain rate by Node $2$ in Figure 7, it can also be decoded reliably by Node $2$ in Figure 7 at the same rate.

Hence we have proved the theorem. ■

The secrecy capacity of the model in Figure 12 is given by Corollary 4. From Corollary 4, we know $R_1 \leq C(\bar{P}_2)$. We next invoke the same technique we used in the proof of Theorem 2 in Appendix B to show $C(\bar{P}_2)$ is also an upper bound on the sum rate. We prove this statement by showing if $R_1 = r_1, R_2 = r_2$ is achievable, then $R_1 = r_1 + r_2$ is also achievable.

Let us construct a message set $\{W_a\}$ which has the same cardinality of the message set $\{W_2\}$. Let part of the secret message be transmitted via $W_a$. The remaining part of the secret message be transmitted via $W_1$. The role of $W_2$ is to be the secret key. Let $W_2$ be taken from the set $\{W_2\}$ according to a uniform distribution. $W_2$ is independent from $W_a$ and $W_1$.

Let $\oplus$ be the modulus addition defined over $\{1, ... \|W_2\|\}$. Node $1$, after decoding $W_2$, transmits $\hat{W}_2 \oplus W_a$ over the public channel. Since the public channel is noiseless with continuous input, it can transmit $\hat{W}_2 \oplus W_a$ with less than $n$ channel uses. Because Node $2$ knows $W_2$, it can recover $W_a$ from $\hat{W}_2 \oplus W_a$ if $W_2 = \hat{W}_2$.

The signal available to the eavesdropper now becomes the output of the wiretap channel $X^n$, and the output of the public link $W_a \oplus W_2$. Then, by the same derivation in (199)-(215), by replacing $Z^n$ with $X^n$, we have:

$$H\left(W_1, W_a | X^n, W_a \oplus \hat{W}_2\right) \tag{142}$$

$$\geq H\left(W_1, W_a\right) - n\varepsilon \tag{143}$$

where $\varepsilon > 0$. $\lim_{n\to\infty} \varepsilon = 0$. Hence the rate of $W_1, W_a$ is the secrecy rate $R_1$. Since $W_a$ is chosen from the message set $\{W_a\}$ according to a uniform distribution, we have $R_1 = r_1 + r_2$.

Due to the symmetry of the channel model, we can prove that $R_2 \leq C(\bar{P}_1)$ and $R_1 + R_2 \leq C(\bar{P}_1)$ in the same fashion.

This completes the proof.

## VI. Conclusion

In this work, we have investigated the merit of using the signals received by the source node, i.e., the feedback, for encoder design on achieving a larger secrecy rate region. In order to answer this question, we studied two models: the Gaussian two-way wiretap channel, and the Gaussian half-duplex two-way relay channel with an untrusted relay. For each model, we derived a computable outer bound for the secrecy capacity region. For the first model, by measuring the gap between the outer bound and the achievable rate region, we find the loss in secrecy rate due to ignoring the feedback signals can be unbounded. Hence the use of feedback can be highly beneficial in this model. For the second model, we find the feedback can be safely ignored if the power of the relay is abundant. In particular, the gap between the achievable rate region and the outer bound is bounded by $0.5$ bit per channel use when the power of the relay goes to $\infty$. It is worth mentioning that the achievable rate region in this case is attained via a time sharing cooperative jamming scheme, which, with its simplicity and near optimum performance, is a viable alternative to an encoding scheme that utilizes feedback signals.

## Appendix A
## Proof of Theorem 1

Let $\varepsilon = \varepsilon_1 + \varepsilon_3$, where $\varepsilon_1$ was defined in (6), and $\varepsilon_3$ was defined in (8). To simplify the notation, we use $M_2'$ to denote $\{M_2, W_2\}$. Then we have:

$$H\left(W_1\right) - n\varepsilon \tag{144}$$

$$\leq H\left(W_1|Z^n\right) - H\left(W_1|Z^n, X_f^n, Y^n, M_2'\right) \tag{145}$$

$$= I\left(W_1; M_2', X_f^n, Y^n|Z^n\right) \tag{146}$$

$$= I\left(W_1; X_f^n|Z^n, Y^n, M_2'\right) + I\left(W_1; M_2', Y^n|Z^n\right) \tag{147}$$

$$= I\left(W_1; M_2', Y^n|Z^n\right) \tag{148}$$

$$\leq I\left(W_1, M_1, Y_f^n; M_2', Y^n|Z^n\right) \tag{149}$$

$$= I\left(W_1, M_1, Y_f^n; M_2', Y^n, Z^n\right) - I\left(W_1, M_1, Y_f^n; Z^n\right) \tag{150}$$

where in (145) follows from (6) and (8). Note that since, in this proof, we are only bounding the rate of $W_1$, we omit $W_2$ from the condition term of (6). (148) is based on the fact that $X_f^n$ is a deterministic function of $Y^{n-1}$ and $M_2'$, as shown in (4).

Then we rewrite the first term in (150) as:

$$I\left(W_1, M_1, Y_f^n; M_2', Y^n, Z^n\right) \tag{151}$$

$$= I\left(W_1, M_1, Y_f^n; Y_n|Z_n, M_2', Y^{n-1}, Z^{n-1}\right) + I\left(W_1, M_1, Y_f^n; Y^{n-1}, Z^n, M_2'\right) \tag{152}$$

For the first term in (152), we have:

$$I\left(W_1, M_1, Y_f^n; Y_n|Z_n, M_2', Y^{n-1}, Z^{n-1}\right) \tag{153}$$

$$= I\left(W_1, M_1, Y_f^n; Y_n|X_{f,n}, Z_n, M_2', Y^{n-1}, Z^{n-1}\right) \tag{154}$$

$$\leq h\left(Y_n|Z_n, X_{f,n}\right) - h\left(Y_n|X_{f,n}, M_2', Y^{n-1}, Z^n, W_1, M_1, Y_f^n\right) \tag{155}$$

$$= h\left(Y_n|Z_n, X_{f,n}\right) - h\left(Y_n|X_{f,n}, X_n, M_2', Y^{n-1}, Z^n, W_1, M_1, Y_f^n\right) \tag{156}$$

$$= h\left(Y_n|Z_n, X_{f,n}\right) - h\left(Y_n|X_{f,n}, X_n, Z_n\right) \tag{157}$$

$$= I\left(X_n; Y_n|Z_n, X_{f,n}\right) \tag{158}$$

In (154), we use the fact that $X_{f,n}$ is a deterministic function of $\{M_2', Y^{n-1}\}$, as shown by (4). In (156), we use the fact that $X_n$ is a deterministic function of $\{W_1, M_1, Y_f^{n-1}\}$, as shown by (3). In (157), we use the fact that

$$Y_n - \{X_{f,n}, X_n, Z_n\} - \{M_2', Y^{n-1}, Z^{n-1}, W_1, M_1, Y_f^n\} \tag{159}$$

is a Markov chain, due to (1) and the channel being memoryless and the fact that encoding functions are causal. In particular, (1) allows us to remove $Y_{f,n}$ from the condition term. Applying this result, we find that (150) is upper bounded by

$$I\left(X_n;Y_n|Z_n,X_{f,n}\right)+I\left(W_1,M_1,Y_f^n;Y^{n-1},M_2'|Z^n\right) \tag{160}$$

The second term in (160) can be rewritten as:

$$I\left(W_1,M_1,Y_f^n;Y^{n-1},M_2'|Z^n\right) \tag{161}$$

$$=I\left(W_1,M_1,Y_f^{n-1};Y^{n-1},M_2'|Z^n\right)+I\left(Y_{f,n};Y^{n-1},M_2'|W_1,M_1,Y_f^{n-1},Z^n\right) \tag{162}$$

The second term in (162) can be upper bounded as:

$$I\left(Y_{f,n};Y^{n-1},M_2'|W_1,M_1,Y_f^{n-1},Z^n\right) \tag{163}$$

$$=I\left(Y_{f,n};Y^{n-1},M_2'|X_n,W_1,M_1,Y_f^{n-1},Z^n\right) \tag{164}$$

$$=h\left(Y_{f,n}|X_n,W_1,M_1,Y_f^{n-1},Z^n\right)-h\left(Y_{f,n}|X_n,W_1,M_1,Y_f^{n-1},Z^n,Y^{n-1},M_2'\right) \tag{165}$$

$$\leq h\left(Y_{f,n}|X_n,Z_n\right)-h\left(Y_{f,n}|X_n,W_1,M_1,Y_f^{n-1},Z^n,Y^{n-1},M_2'\right) \tag{166}$$

$$=h\left(Y_{f,n}|X_n,Z_n\right)-h\left(Y_{f,n}|X_{f,n},X_n,Z_n,W_1,M_1,Y_f^{n-1},Z^{n-1},Y^{n-1},M_2'\right) \tag{167}$$

$$=h\left(Y_{f,n}|X_n,Z_n\right)-h\left(Y_{f,n}|X_{f,n},X_n,Z_n\right) \tag{168}$$

$$=I(X_{f,n};Y_{f,n}|X_n,Z_n) \tag{169}$$

In (164), we use the fact that $X_n$ is a deterministic function of $\{W_1,M_1,Y_f^{n-1}\}$, as shown by (3). In (167), we use the fact that $X_{f,n}$ is a deterministic function of $M_2',Y^{n-1}$, as shown by (4). In (168), we use the fact that

$$Y_{f,n}-\{X_{f,n},X_n,Z_n\}-\{W_1,M_1,Y_f^{n-1},Z^{n-1},Y^{n-1},M_2'\} \tag{170}$$

is a Markov chain. This is because the encoding functions are causal and the channel is memoryless.

Applying this result, we find that that (160) is now upper bounded by

$$I\left(X_n;Y_n|Z_n,X_{f,n}\right)+I(X_{f,n};Y_{f,n}|X_n,Z_n)+I\left(W_1,M_1,Y_f^{n-1};Y^{n-1},M_2'|Z^n\right) \tag{171}$$

The last term in (171) can be rewritten as

$$I\left(W_1,M_1,Y_f^{n-1};Y^{n-1},M_2'|Z^{n-1}\right)+I\left(W_1,M_1,Y_f^{n-1};Z_n|Y^{n-1},M_2',Z^{n-1}\right)$$
$$-I\left(W_1,M_1,Y_f^{n-1};Z_n|Z^{n-1}\right) \tag{172}$$

The second term and the last term in (172) can be upper bounded together:

$$
\begin{aligned}
& I\left(W_1, M_1, Y_f^{n-1}; Z_n | M_2', Y^{n-1}, Z^{n-1}\right) - I\left(W_1, M_1, Y_f^{n-1}; Z_n | Z^{n-1}\right) && (173)\\
=& - I\left(Z_n; M_2', Y^{n-1} | Z^{n-1}\right) - h\left(Z_n | W_1, M_1, Y_f^{n-1}, M_2', Y^{n-1}, Z^{n-1}\right) \\
& + h\left(Z_n | Z^{n-1}, W_1, M_1, Y_f^{n-1}\right) && (174)\\
\leq& - h\left(Z_n | W_1, M_1, Y_f^{n-1}, M_2', Y^{n-1}, Z^{n-1}\right) + h\left(Z_n | Z^{n-1}, W_1, M_1, Y_f^{n-1}\right) && (175)\\
=& - h\left(Z_n | X_n, X_{f,n}, W_1, M_1, Y_f^{n-1}, M_2', Y^{n-1}, Z^{n-1}\right) + h\left(Z_n | X_n, Z^{n-1}, W_1, M_1, Y_f^{n-1}\right) && (176)\\
\leq& - h\left(Z_n | X_n, X_{f,n}, W_1, M_1, Y_f^{n-1}, M_2', Y^{n-1}, Z^{n-1}\right) + h\left(Z_n | X_n\right) && (177)\\
=& - h\left(Z_n | X_n, X_{f,n}\right) + h\left(Z_n | X_n\right) && (178)\\
=& I(X_{f,n}; Z_n | X_n) && (179)
\end{aligned}
$$

In (176), we use the fact that $X_n$ is a deterministic function of $\{W_1, M_1, Y_f^{n-1}\}$, and $X_{f,n}$ is a deterministic function of $\{M_2', Y^{n-1}\}$. In (178), we use the fact that

$$
Z_n - \{X_n, X_{f,n}\} - \{W_1, M_1, Y_f^{n-1}, M_2', Y^{n-1}, Z^{n-1}\} \tag{180}
$$

is a Markov chain. This is due to the fact that the channel is memoryless and the encoding functions (3) and (4) are causal.

Applying this result to (172), we find that that (171) is now upper bounded by:

$$
I\left(X_n; Y_n | X_{f,n}, Z_n\right) + I(X_{f,n}; Y_{f,n}, Z_n | X_n) + I\left(W_1, M_1, Y_f^{n-1}; Y^{n-1}, M_2' | Z^{n-1}\right) \tag{181}
$$

Hence we have shown that

$$
\begin{aligned}
H(W_1) - n\varepsilon \leq& I\left(W_1, M_1, Y_f^n; Y^n, M_2' | Z^n\right) \\
\leq& I\left(X_n; Y_n | X_{f,n}, Z_n\right) + I\left(X_{f,n}; Y_{f,n}, Z_n | X_n\right) \\
& + I\left(W_1, M_1, Y_f^{n-1}; Y^{n-1}, M_2' | Z^{n-1}\right) && (182)
\end{aligned}
$$

Applying this result repeatedly for $n-1, n-2, ..., 1$, we have

$$
\begin{aligned}
& \frac{1}{n} H(W_1) - \varepsilon && (183)\\
\leq& \frac{1}{n} \sum_{i=1}^{n} (I(X_i; Y_i | X_{f,i}, Z_i) + I(X_{f,i}; Y_{f,i}, Z_i | X_i)) && (184)
\end{aligned}
$$

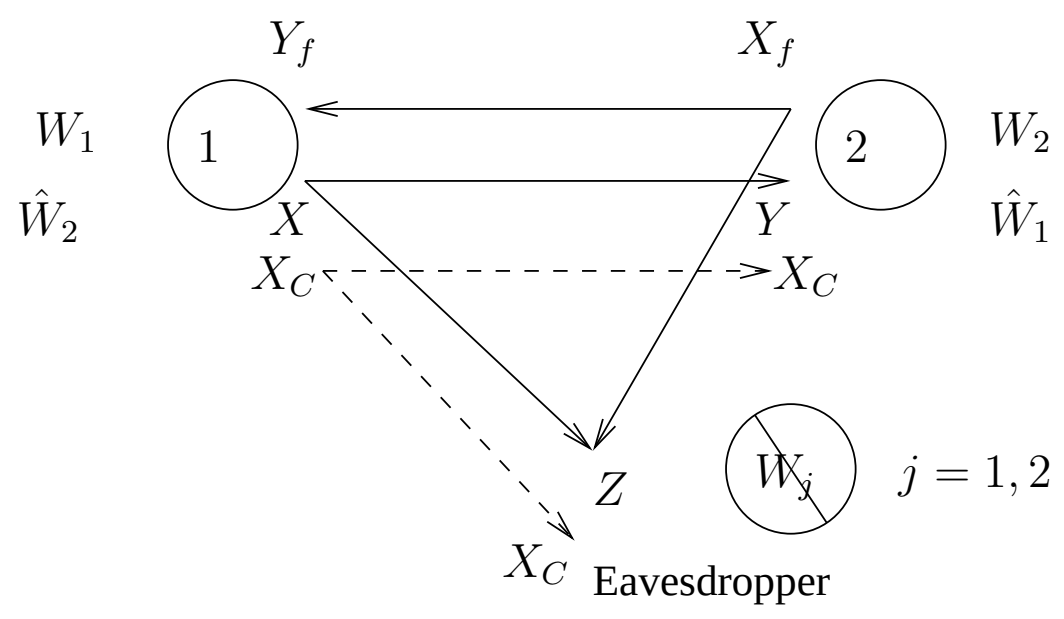

Fig. 13. Two-way wiretap channel with a public noiseless forward link

Define $Q$ as a random variable that is uniformly distributed over $\{1, 2, ..., n\}$. Define $X = X_Q, Y = Y_Q, Z = Z_Q, X_f = X_{f,Q}, Y_f = Y_{f,Q}$. Then the right hand side of (184) equals:

$$I(X; Y|Z, X_f, Q) + I(X_f; Y_f, Z|X, Q) \tag{185}$$

$$\leq I\left(X; Y|Z, X_f\right) + I(X_f; Y_f, Z|X) \tag{186}$$

where we use the fact that $Y - \{Z, X_f, X\} - Q$ is a Markov chain and $\{Y_f, Z\} - \{X, X_f\} - Q$ is a Markov chain. Applying this result in (184) and let $n \to \infty$, we obtained the upper bound in the theorem.

## Appendix B
## Proof of Theorem 2

Equation (47) follows from removing the eavesdropper and applying the bounds of two-way channel from [1]. Equation (48) can be derived similarly thanks to the symmetry of the channel model.

We next derive (49). We focus on the first term inside the minimum in (49). The second term can be derived similarly thanks to the symmetry of the channel model.

First we add a public noiseless broadcast channel to the channel in Figure 1. The new channel model is shown in Figure 13. The broadcast channel takes the input from Node $1$. Its outputs are received by Node $2$ and the eavesdropper. Since the channel is noiseless, the outputs equal the input, and is denoted by $X_C$. $X_C$ is continuous. The introduction of the public noiseless broadcast channel certainly does not decrease the secrecy capacity region. Hence, to upper bound the secrecy capacity region of the original channel, we can consider this new model instead. We next apply Theorem 1 to this channel, which says $R_1$ is bounded by

$$I(X, X_C; Y, X_C|Z, X_C, X_f) + I(X_f; Y_f, Z, X_C|X, X_C) \tag{187}$$

The first term in (187) is upper bounded by:

$$\begin{aligned} I\left(X, X_C; Y, X_C|Z, X_C, X_f\right) =& I\left(X; Y|Z, X_C, X_f\right) & (188)\\ =& h\left(Y|Z, X_C, X_f\right) - h\left(Y|Z, X, X_C, X_f\right) & (189)\\ \leq& h\left(Y|Z, X_f\right) - h\left(Y|Z, X, X_C, X_f\right) & (190)\\ =& h\left(Y|Z, X_f\right) - h\left(Y|Z, X, X_f\right) & (191)\\ =& I\left(X; Y|Z, X_f\right) & (192)\end{aligned}$$

In (191) we use the fact that $Y - \{Z, X, X_f\} - X_C$ is a Markov chain.

The second term is (187) is upper bounded by:

$$\begin{aligned} & I\left(X_f; Y_f, Z, X_C|X, X_C\right) & (193)\\ =& I\left(X_f; Y_f, Z|X, X_C\right) & (194)\\ \leq& h\left(Y_f, Z|X\right) - h\left(Y_f, Z|X, X_f, X_C\right) & (195)\\ =& h\left(Y_f, Z|X\right) - h\left(Y_f, Z|X, X_f\right) & (196)\\ =& I\left(X_f; Y_f, Z|X\right) & (197)\end{aligned}$$

In (196) we use the fact that $\{Y_f, Z\} - \{X, X_f\} - X_C$ is a Markov chain.

Hence (187) is upper bounded by

$$I(X; Y|Z, X_f) + I(X_f; Y_f, Z|X) \tag{198}$$

This means introducing a public noiseless forward channel brings no change in the expression of the upper bound of $R_1$.

We next prove (198) is also an upper bound on $R_1 + R_2$. This is done by showing if $R_1 = r_1, R_2 = r_2$ is achievable, then $R_1 = r_1 + r_2$ is also achievable.

Construct a message set $\{W_a\}$ which has the same cardinality of the message set $\{W_2\}$. Let part of the secret message be transmitted via $W_a$. The remaining part of the secret message is transmitted via $W_1$. The role of $W_2$ is to serve as a secret key. Let $W_2$ be taken from the set $\{W_2\}$ according to a uniform distribution. $W_2$ is independent from $W_a$ and $W_1$.

Let $\oplus$ be the modulus addition defined over $\{1, ... \|W_2\|\}$, where $\|W_2\|$ is the cardinality of the set $\{W_2\}$. Recall that $\hat{W}_2$ denote the result obtained by Node $1$ when it tries to decode $W_2$. We let Node $1$ transmit $\hat{W}_2 \oplus W_a$ over the public channel. Since the public channel is noiseless with continuous input, it can transmit $\hat{W}_2 \oplus W_a$ with a single channel use. Because Node $2$ knows $W_2$, it can recover $W_a$ from $\hat{W}_2 \oplus W_a$ when $W_2 = \hat{W}_2$.

The signal available to the eavesdropper now becomes the output of the wiretap channel $Z^n$, and the output of the public link, which is $W_a \oplus \hat{W}_2$. Conditioned on these signals, the equivocation of $W_1, W_a$ can be computed as:

$$H\left(W_1, W_a | Z^n, W_a \oplus \hat{W}_2\right) \tag{199}$$

$$\geq H\left(W_1, W_a | Z^n, W_a \oplus \hat{W}_2, W_a \oplus W_2\right) \tag{200}$$

$$= H\left(W_1, W_a, W_a \oplus \hat{W}_2 | Z^n, W_a \oplus W_2\right) - H\left(W_a \oplus \hat{W}_2 | Z^n, W_a \oplus W_2\right) \tag{201}$$

$$= H\left(W_1, W_a | Z^n, W_a \oplus W_2\right) + H\left(W_a \oplus \hat{W}_2 | W_1, W_a, Z^n, W_a \oplus W_2\right)$$

$$- H\left(W_a \oplus \hat{W}_2 | Z^n, W_a \oplus W_2\right) \tag{202}$$

$$\geq H\left(W_1, W_a | Z^n, W_a \oplus W_2\right) - H\left(W_a \oplus \hat{W}_2 | Z^n, W_a \oplus W_2\right) \tag{203}$$

$$\geq H\left(W_1, W_a | Z^n, W_a \oplus W_2\right) - H\left(W_a \oplus \hat{W}_2 | W_a, Z^n, W_a \oplus W_2\right) \tag{204}$$

$$= H\left(W_1, W_a | Z^n, W_a \oplus W_2\right) - H\left(\hat{W}_2 | W_a, Z^n, W_2\right) \tag{205}$$

$$\geq H\left(W_1, W_a | Z^n, W_a \oplus W_2\right) - H\left(\hat{W}_2 | W_2\right) \tag{206}$$

$$\geq H\left(W_1, W_a | Z^n, W_a \oplus W_2\right) - n\varepsilon \tag{207}$$

In (207) we use the fact that $W_2$ can be reliably decoded by Node $1$. Hence (207) follows from Fano's inequality.

The first term in (207) can be bounded as follows:

$$H\left(W_1, W_a | Z^n, W_a \oplus W_2\right) \tag{208}$$

$$= H\left(W_a | Z^n, W_a \oplus W_2\right) + H\left(W_1 | Z^n, W_a, W_a \oplus W_2\right) \tag{209}$$

$$= H\left(W_a | Z^n, W_a \oplus W_2\right) + H\left(W_1 | Z^n, W_a, W_2\right) \tag{210}$$

$$= H\left(W_a | W_a \oplus W_2\right) + H\left(W_1 | Z^n, W_a, W_2\right) \tag{211}$$

$$= H\left(W_a | W_a \oplus W_2\right) + H\left(W_1 | Z^n, W_2\right) \tag{212}$$

$$=H(W_a)+H(W_1|Z^n,W_2) \tag{213}$$

$$\geq H(W_a)+H(W_1)-n\varepsilon \tag{214}$$

$$\geq H(W_1,W_a)-n\varepsilon \tag{215}$$

Equation (211) is due to the fact that $Z^n$ is independent from $W_a, W_2$, which leads to:

$$I(W_a;Z^n|W_a\oplus W_2)\leq I(W_a,W_a\oplus W_2;Z^n)=I(W_a,W_2;Z^n)=0 \tag{216}$$

Equation (212) follows from the fact that $W_a$ is independent from $Z^n$,$W_1$,$W_2$. Equation (214) follows from the fact that collective secrecy implies one message is secure even if the other message is revealed to the eavesdropper [12].

The argument above shows the rate of $W_1, W_a$ is the secrecy rate $R_1$. Since $W_a$ is chosen from the message set $\{W_a\}$ according to a uniform distribution, we have $R_1 = r_1 + r_2$.

Therefore $R_1 + R_2$ is upper bounded by (198).

Hence we have proved the theorem.

## Appendix C
## Proof of Theorem 6

We prove $R_1 = R_1^*, R_2 = 0$ is achievable. The achievability of $R_1 = 0, R_2 = R_2^*$ can be proved similarly due to the symmetry of the channel model.

The communication is divided into two phases:

1) The first phase lasts $n$ channel uses. During it, Node $2$ sends a key $K$ to Node $1$. At the same time, Node $1$ performs cooperative jamming by transmitting an i.i.d. Gaussian noise sequence with power $P$.
2) The second phase lasts $\bar{n}$ channel uses, during which Node $1$ encrypts the confidential message $W$ with $K$, and sends the result back to Node $2$. At the same time, Node $2$ performs cooperative jamming by transmitting an i.i.d. Gaussian noise sequence with power $P_r$.

Let $\alpha = n/(n+\bar{n})$ be the time sharing factor of the first phase. $0 \leq \alpha \leq 1$ and $\alpha$ is a constant.

The following notation is used in the remainder of the proof: $\bar{x}$ denotes any signal $x$ which is related to the second phase. Otherwise, the signal is related to the first phase. With this notation, the signals received by the eavesdropper during the two phases are given by:

$$Z^n = \sqrt{h_1}X^n + \sqrt{h_2}X_f^n + N_2^n \tag{217}$$

$$\bar{Z}^{\bar{n}} = \sqrt{h_1}\bar{X}^{\bar{n}} + \sqrt{h_1}\bar{X}_f^{\bar{n}} + \bar{N}_2^{\bar{n}} \tag{218}$$

The codebooks used by Node $1$ and $2$ are denoted by $\mathcal{C}_1$ and $\mathcal{C}_2$ respectively and are generated in the following way:

$\mathcal{C}_2$ is composed of i.i.d. sequences sampled from the Gaussian distribution $\mathcal{N}(0, P_r)$. The codebook is then randomly binned into several bins. The size of the codebook depends on the number of bins needed to represent the key $K$ and the size of the bin necessary to confuse the eavesdropper. Specifically, the size of the bin is chosen to be

$$2^{\lfloor n(C\left(\frac{h_2P_r}{h_1P+1}\right)-\epsilon)\rfloor} \tag{219}$$

where $\lfloor x \rfloor$ denotes the largest integer smaller or equal to $x$, $\epsilon > 0$ and $\lim_{n\to\infty}\epsilon = 0$.

Let $R_K$ be the rate of the secret key. Then there are $2^{nR_K}$ bins. $R_K$ is given by:

$$0 < R_K = \frac{1}{n}H\left(K|\mathcal{C}_1,\mathcal{C}_2\right) < \min\left\{\left[C\left(P_r\right) - C\left(\frac{h_2P_r}{h_1P+1}\right)\right]^+, C\left(\frac{h_1P}{h_2P_r+1}\right)\right\} \tag{220}$$

Observe that the key rate is chosen to be smaller than $\left[C\left(P_r\right) - C\left(\frac{h_2P_r}{h_1P+1}\right)\right]^+$ to keep the key $K$ secret from the eavesdropper. As will be shown later, the key is used to compensate the rate loss of the forward channel needed to confuse to eavesdropper. Hence, the rate of the key is chosen not to exceed this rate loss, which leads to the term $C\left(\frac{h_1P}{h_2P_r+1}\right)$ in (220).

$\mathcal{C}_1$ is composed of $2^{nR_K}$ codebooks. Each codebook is composed of i.i.d. sequences sampled from the Gaussian distribution $\mathcal{N}(0, P)$, and is composed of $2^{\bar{n}C(P)}$ i.i.d. Gaussian sequences. The sequences of each codebook are randomly binned into several bins. The size of each bin is chosen to be:

$$2^{\lfloor(\bar{n}C\left(\frac{h_1P}{h_2P_r+1}\right)-nR_K-\bar{n}\epsilon_1)\rfloor} \tag{221}$$

where $\epsilon_1 > 0$ and $\lim_{n\to\infty}\epsilon_1 = 0$.

During the first phase, Node $2$ generates a secret key $K$ according to a uniform distribution over $\{1, ..., 2^{nR_K}\}$ and selects the bin from $\mathcal{C}_2$ according to $K$. Then it chooses a codeword from this bin according to a uniform distribution and transmits it to Node $1$.

Since Node $1$ is transmitting an i.i.d. Gaussian noise sequence during the first phase, the channel model in this phase is equivalent to the Gaussian wiretap channel [29], which uses the same codebook and encoding scheme as we do here. Reference [29] proves that, by doing so, $K$ is kept secret from the eavesdropper and can be reliably decoded by Node $1$. That is:

$$\frac{1}{n} I\left(K; Z^n | \mathcal{C}_1, \mathcal{C}_2\right) \leq \varepsilon \tag{222}$$

$$\lim_{n\to\infty} E[\Pr(\hat{K} \neq K | \mathcal{C}_1, \mathcal{C}_2)] = 0 \tag{223}$$

where $\varepsilon \geq 0$, $\lim_{n\to\infty} \varepsilon = 0$.

Let $\hat{K}$ be the estimate of $K$ Node $1$ decodes from its received signal. Node $1$ computes its transmitted signals as follows: It first chooses the codebook according to the key $\hat{K}$ it decoded from the first phase. Then, it chooses the bin from the codebook according to the secret message $W$. Finally, it chooses the transmitted codeword from this bin according to a uniform distribution.

If $\hat{K} = K$, then Node $2$ knows the sub-codebook used by Node $1$. The sub-codebook is composed of i.i.d. Gaussian sequences and its rate is within the AWGN channel capacity between Node $1$ and Node $2$. This observation, along with (223), leads to the following fact:

$$\lim_{\bar{n}\to\infty} E[\Pr(\hat{W} \neq W | \mathcal{C}_1, \mathcal{C}_2)] = 0 \tag{224}$$

We next bound the equivocation

$$H\left(W | Z^n, \bar{Z}^{\bar{n}}, \mathcal{C}_1, \mathcal{C}_2\right) \tag{225}$$

It is understood that $\mathcal{C}_1, \mathcal{C}_2$ is always on the condition term. Hence, we omit it in the sequel to simplify the notation and reinstate it only when necessary.

The equivocation rate is then bounded as follows:

$$\begin{aligned}
& H\left(W | Z^n, \bar{Z}^{\bar{n}}\right) \\
=& H\left(\bar{X}^{\bar{n}}, W | Z^n, \bar{Z}^{\bar{n}}\right) - H\left(\bar{X}^{\bar{n}} | W, Z^n, \bar{Z}^{\bar{n}}\right) && (226) \\
\geq& H\left(\bar{X}^{\bar{n}}, W | Z^n, \bar{Z}^{\bar{n}}\right) - \bar{n}\varepsilon && (227) \\
=& H\left(W | Z^n, \bar{Z}^{\bar{n}}, \bar{X}^{\bar{n}}\right) + H\left(\bar{X}^{\bar{n}} | Z^n, \bar{Z}^{\bar{n}}\right) - \bar{n}\varepsilon && (228) \\
=& H\left(\bar{X}^{\bar{n}} | Z^n, \bar{Z}^{\bar{n}}\right) - \bar{n}\varepsilon && (229) \\
=& H\left(\bar{X}^{\bar{n}} | Z^n, \bar{Z}^{\bar{n}}\right) - H\left(\bar{X}^{\bar{n}}\right) + H\left(\bar{X}^{\bar{n}}\right) - \bar{n}\varepsilon && (230)
\end{aligned}$$

$$=H\left(\bar{X}^{\bar{n}}\right)-I\left(\bar{X}^{\bar{n}};Z^{n},\bar{Z}^{\bar{n}}\right)-\bar{n}\varepsilon \tag{231}$$

$$=H\left(\bar{X}^{\bar{n}}\right)-I\left(\bar{X}^{\bar{n}};\bar{Z}^{\bar{n}}\right)-I\left(\bar{X}^{\bar{n}};Z^{n}|\bar{Z}^{\bar{n}}\right)-\bar{n}\varepsilon \tag{232}$$

Here (227) follows from the fact that given $W$, the number of possible $\bar{X}^{\bar{n}}$ equals the cardinality of the bin that corresponds to $W$ from all the $2^{nR_K}$ codebooks, which is $2^{\bar{n}\left(C\left(\frac{h_1P}{h_2P_r+1}\right)-\epsilon_1\right)}$. Note that these candidates of $\bar{X}^{\bar{n}}$ form a Gaussian codebook by itself with a rate of $C\left(\frac{h_1P}{h_2P_r+1}\right)-\epsilon_1$. Since Node $2$ is transmitting i.i.d. Gaussian noise, the channel between Node $1$ and the eavesdropper is an AWGN channel whose capacity is $C\left(\frac{h_1P}{h_2P_r+1}\right)$. Therefore, given $W$, the eavesdropper can determine $\bar{X}^{\bar{n}}$ from $\bar{Z}^{\bar{n}}$ using joint typical decoding. (227) then follows by applying Fano's inequality.

Equation (229) follows since $W$ is a deterministic function of $\bar{X}^{\bar{n}}$.

The third term in (232) can then be bounded as follows:

$$I\left(\bar{X}^{\bar{n}};Z^{n}|\bar{Z}^{\bar{n}}\right)=h\left(Z^{n}|\bar{Z}^{\bar{n}}\right)-h\left(Z^{n}|\bar{Z}^{\bar{n}},\bar{X}^{\bar{n}}\right) \tag{233}$$

$$=h\left(Z^{n}|\bar{Z}^{\bar{n}}\right)-h\left(Z^{n}|\bar{X}_f^{\bar{n}}+\bar{N}_2^{\bar{n}},\bar{X}^{\bar{n}}\right) \tag{234}$$

$$=h\left(Z^{n}|\bar{Z}^{\bar{n}}\right)-h\left(Z^{n}|\bar{X}^{\bar{n}}\right) \tag{235}$$

$$\leq h\left(Z^{n}|\bar{Z}^{\bar{n}}\right)-h\left(Z^{n}|\bar{X}^{\bar{n}},K\right) \tag{236}$$

$$=h\left(Z^{n}|\bar{Z}^{\bar{n}}\right)-h\left(Z^{n}|K\right) \tag{237}$$

$$=h\left(Z^{n}|\bar{Z}^{\bar{n}}\right)-h\left(Z^{n}\right)-h\left(Z^{n}|K\right)+h\left(Z^{n}\right) \tag{238}$$

$$=I\left(Z^{n};K\right)-I\left(Z^{n};\bar{Z}^{\bar{n}}\right) \tag{239}$$

$$\leq I\left(Z^{n};K\right)\leq n\varepsilon_2 \tag{240}$$

Equation (235) is because $\bar{X}_f^{\bar{n}}+\bar{N}_2^{\bar{n}}$ is a sequence of i.i.d. Gaussian noise, which is independent from $Z^n$ and $\bar{X}^{\bar{n}}$.

Equation (237) follows from the fact that $Z^n-K-\bar{X}^{\bar{n}}$ is a Markov chain. Equation (240) follows from (222).

Substituting (240) into (232), we have

$$H\left(W|Z^{n},\bar{Z}^{\bar{n}}\right) \tag{241}$$

$$\geq H\left(\bar{X}^{\bar{n}}\right)-I\left(\bar{X}^{\bar{n}};\bar{Z}^{\bar{n}}\right)-\left(\bar{n}\varepsilon+n\varepsilon_2\right) \tag{242}$$

The second term in (242) can be bounded as follows. For this purpose, we reinstate the $\mathcal{C}_1, \mathcal{C}_2$ on the condition term:

$$
\begin{aligned}
& I\left(\bar{X}^{\bar{n}}; \bar{Z}^{\bar{n}} | \mathcal{C}_1, \mathcal{C}_2\right) && (243)\\
\leq & h\left(\bar{Z}^{\bar{n}}\right) - h\left(\bar{Z}^{\bar{n}} | \bar{X}^{\bar{n}}, \mathcal{C}_1, \mathcal{C}_2\right) && (244)\\
= & h\left(\bar{Z}^{\bar{n}}\right) - h\left(\bar{Z}^{\bar{n}} | \bar{X}^{\bar{n}}, \mathcal{C}_1, \mathcal{C}_2\right) && (245)\\
= & h\left(\bar{Z}^{\bar{n}}\right) - h\left(\bar{Z}^{\bar{n}} | \bar{X}^{\bar{n}}\right) && (246)\\
= & I\left(\bar{X}^{\bar{n}}; \bar{Z}^{\bar{n}}\right) && (247)\\
= & \bar{n} I\left(\bar{X}; \bar{Z}\right) && (248)
\end{aligned}
$$

Equation (246) follows from the fact that given $\bar{X}^{\bar{n}}$, $\bar{Z}^{\bar{n}}$ only depends on the jamming signal and channel noise. Therefore, we can drop codebooks $\mathcal{C}_1, \mathcal{C}_2$ from the conditioning term. Equation (248) follows from the fact that Node $2$ transmits i.i.d. Gaussian noise during the second phase, and the code book used by Node $1$ is composed of i.i.d. Gaussian sequences.

Since

$$
I\left(\bar{X}; \bar{Z}\right) = C\left(\frac{h_1 P}{h_2 P_r + 1}\right) \tag{249}
$$

$$
H\left(\bar{X}^{\bar{n}} | \mathcal{C}_1, \mathcal{C}_2\right) = n R_K + \bar{n} C\left(P\right) \tag{250}
$$

we have

$$
\begin{aligned}
& H\left(W | Z^n, \bar{Z}^{\bar{n}}, \mathcal{C}_1, \mathcal{C}_2\right) && (251)\\
= & \left(n R_K + \bar{n} C\left(P\right)\right) - \bar{n} C\left(\frac{h_1 P}{h_2 P_r + 1}\right) - \left(\bar{n}\varepsilon + n\varepsilon_2\right) && (252)\\
\geq & H(W | \mathcal{C}_1, \mathcal{C}_2) - \left(\bar{n}(\varepsilon + \epsilon_1) + n\varepsilon_2\right) && (253)
\end{aligned}
$$

Therefore $0 \leq I(W; Z^n, \bar{Z}^{\bar{n}} | \mathcal{C}_1, \mathcal{C}_2) < \left(\bar{n}(\varepsilon + \epsilon_1) + n\varepsilon_2\right)$. This, along with (224), gives us:

$$
\lim_{n, \bar{n} \to \infty} \frac{1}{n + \bar{n}} I(W; Z^n, \bar{Z}^{\bar{n}} | \mathcal{C}_1, \mathcal{C}_2) + E[\Pr(\hat{W} \neq W) | \mathcal{C}_1, \mathcal{C}_2] = 0 \tag{254}
$$

From the linearity of expectation and non-negativity of mutual information and probability, we see that there must exists codebooks $\mathcal{C}_1 = \mathcal{C}_1^*, \mathcal{C}_2 = \mathcal{C}_2^*$ such that both terms on the left hand side of (254) go to $0$ as $n, \bar{n} \to \infty$. This observation, along with that fact that $n + \bar{n}$ channel uses are involved, proves that the secrecy rate pair $(R_1^*, 0)$ is achievable.

Hence we have proved the theorem.

# APPENDIX D
## PROOF OF THEOREM 7

Since received signals are not used to compute transmitting signals at Node $j$, $j = 1, 2$, we let $\alpha = 1$ in Theorem 6. In this case, when $P = kP_r$, $R_j^*$ becomes:

$$R_1^* = C(P) - C\left(\frac{h_1}{h_2 k + 1/P}\right) \tag{255}$$

$$R_2^* = C(kP) - C\left(\frac{h_2 k}{h_1 + 1/P}\right) \tag{256}$$

The sum rate bound given by Theorem 5 is upper bounded by:

$$\min\left\{C\left(\frac{P}{h_1 P + 1}\right) + C\left((h_2+1)kP\right), C\left(\frac{kP}{h_2 kP + 1}\right) + C\left((h_1+1)P\right)\right\} \tag{257}$$

To prove Theorem 7, it is sufficient to show both $R_1^*$ and $R_2^*$ are within constant gaps of (257), as we show below:

$$C\left(\frac{P}{h_1 P + 1}\right) + C\left((h_2+1)kP\right) - R_1^* \tag{258}$$

$$= C\left(\frac{P}{h_1 P + 1}\right) + C\left((h_2+1)kP\right) - C(P) + C\left(\frac{h_1}{h_2 k + 1/P}\right) \tag{259}$$

$$\leq C\left(\frac{P}{h_1 P + 1}\right) + C\left((h_2+1)kP\right) - C(P) + C\left(\frac{h_1}{h_2 k}\right) \tag{260}$$

$$\leq C\left(\frac{1}{h_1}\right) + C\left((h_2+1)kP\right) - C(P) + C\left(\frac{h_1}{h_2 k}\right) \tag{261}$$

$$= C\left(\frac{1}{h_1}\right) + \frac{1}{2}\log_2\left(\frac{1 + (h_2+1)kP}{1+P}\right) + C\left(\frac{h_1}{h_2 k}\right) \tag{262}$$

$$\leq C\left(\frac{1}{h_1}\right) + \frac{1}{2}\log_2\left(\frac{1 + \max\{1, (h_2+1)k\}P}{1+P}\right) + C\left(\frac{h_1}{h_2 k}\right) \tag{263}$$

$$\leq C\left(\frac{1}{h_1}\right) + \frac{1}{2}\log_2\left(\max\{1, (h_2+1)k\}\right) + C\left(\frac{h_1}{h_2 k}\right) \tag{264}$$

For $R_2^*$, we have:

$$C\left(\frac{P}{h_1 P + 1}\right) + C\left((h_2+1)kP\right) - R_2^* \tag{265}$$

$$= C\left(\frac{P}{h_1 P + 1}\right) + C\left((h_2+1)kP\right) - C(kP) + C\left(\frac{h_2 k}{h_1 + 1/P}\right) \tag{266}$$

$$\leq C\left(\frac{P}{h_1 P + 1}\right) + C\left((h_2+1)kP\right) - C(kP) + C\left(\frac{h_2 k}{h_1}\right) \tag{267}$$

$$\leq C\left(\frac{1}{h_1}\right) + C\left(\left(h_2+1\right)kP\right) - C\left(kP\right) + C\left(\frac{h_2 k}{h_1}\right) \tag{268}$$

$$= C\left(\frac{1}{h_1}\right) + \frac{1}{2}\log_2\left(\frac{1+\left(h_2+1\right)kP}{1+kP}\right) + C\left(\frac{h_2 k}{h_1}\right) \tag{269}$$

$$\leq C\left(\frac{1}{h_1}\right) + \frac{1}{2}\log_2\left(h_2+1\right) + C\left(\frac{h_2 k}{h_1}\right) \tag{270}$$

Hence we have proved the Theorem.

## Appendix E
## Proof of Theorem 8

To prove this theorem, we only need show that it is possible to achievable a secrecy rate for Node $1$ that exceeds the upper bound given by Theorem 4. Consider the case when $h_1 = h_2 = 1$. Then by evaluating (57) at $\sigma^2 = 0$ and $\sigma^2 \to \infty$ with $\rho = \eta = 0$, we find the secrecy rate $R_1$ is bounded by

$$\min\{C(P), C(P_r) + 0.5\} \tag{271}$$

when $Y_f$ is ignored by Node $1$. Choose $P_r$ and $P$ such that

$$C(P_r) + 0.5 < 0.4C(P) \tag{272}$$

For this power configuration, from (271), we observe that $R_1$ is upper bounded by $0.4C(P)$.

Let the $\alpha$ in Theorem 6 be $0.5$. $R_1^*$ then becomes:

$$0.5C\left(P\right) - 0.5\left[C\left(\frac{P}{P_r+1}\right) - C\left(P_r\right) + C\left(\frac{P_r}{P+1}\right)\right]^+ \tag{273}$$

A sufficient condition for $R_1^* = 0.5C(P)$ is that

$$C\left(\frac{P}{P_r+1}\right) + C\left(\frac{P_r}{P+1}\right) > C\left(P_r\right) \tag{274}$$

It can be verified that this condition is equivalent to

$$\frac{\left(\frac{P}{P_r+1}+1\right)^2}{P+1} > 1 \tag{275}$$

A sufficient condition for it to hold is:

$$\frac{\left(\frac{P}{P_r+1}+1\right)^2}{\left(\sqrt{P}+1\right)^2} > 1 \tag{276}$$

which means

$$\sqrt{P} > P_r + 1 \tag{277}$$

Choose $P_r = P^{1/4}$. For sufficiently large $P$, both (272) and (277) can be fulfilled. In this case, the achievable rate is $0.5C(P)$, which is greater than the upper bound $0.4C(P)$. The difference is $0.1C(P)$, which is not a bounded function of $P$. Hence we have proved the theorem.

## APPENDIX F
## PROOF OF THEOREM 12

The upper bound $I(X;Y)$ follows from removing the eavesdropper and applying the upper bound for two-way channel from [1]. Hence we only need to prove the second term inside the minimum.

Let $\varepsilon = \varepsilon_4 + \varepsilon_5$, where $\varepsilon_4, \varepsilon_5$ were defined in (111) and (112). Then we have:

$$H(W) - n\varepsilon \tag{278}$$

$$\leq H(W|Z^n) - H\left(W|Z^n, X_f^n, Y^n, M_2\right) \tag{279}$$

$$= I\left(W; M_2, X_f^n, Y^n|Z^n\right) \tag{280}$$

$$= I\left(W; X_f^n|Z^n, Y^n, M_2\right) + I\left(W; M_2, Y^n|Z^n\right) \tag{281}$$

$$= I\left(W; M_2, Y^n|Z^n\right) \tag{282}$$

$$\leq I\left(W, M_1, Y_f^n; M_2, Y^n|Z^n\right) \tag{283}$$

$$= I\left(W, M_1, Y_f^n; M_2, Y^n, Z^n\right) - I\left(W, M_1, Y_f^n; Z^n\right) \tag{284}$$

In (279) we use (111) and (112). In (282) we use the fact that $X_f^n$ is a deterministic function of $Y^{n-1}$ and $M_2$, as shown in (110).

For the first term in (284), we have:

$$I\left(W, M_1, Y_f^n; M_2, Y^n, Z^n\right) \tag{285}$$

$$= I\left(W, M_1, Y_f^n; Y_n|Z_n, M_2, Y^{n-1}, Z^{n-1}\right) + I\left(W, M_1, Y_f^n; Z_n|M_2, Y^{n-1}, Z^{n-1}\right)$$
$$+ I\left(W, M_1, Y_f^n; Y^{n-1}, Z^{n-1}, M_2\right) \tag{286}$$

For the first term in (286), we have:

$$I\left(W, M_1, Y_f^n; Y_n|Z_n, M_2, Y^{n-1}, Z^{n-1}\right) \tag{287}$$

$$\leq h\left(Y_n|Z_n\right) - h\left(Y_n|M_2, Y^{n-1}, Z^n, W, M_1, Y_f^n\right) \tag{288}$$

$$= h\left(Y_n|Z_n\right) - h\left(Y_n|X_n, M_2, Y^{n-1}, Z^n, W, M_1, Y_f^n\right) \tag{289}$$

$$= h\left(Y_n|Z_n\right) - h\left(Y_n|X_n, Z_n\right) \tag{290}$$

$$= I\left(X_n; Y_n|Z_n\right) \tag{291}$$

In (289), we use the fact that $X_n$ is a deterministic function of $W, M_1, Y_f^n$. In (290), we use the fact that $Y_n - \{X_n, Z_n\} - \{M_2, Y^{n-1}, Z^{n-1}, W, M_1, Y_f^n\}$, since the channel is memoryless and encoding functions are causal.

Applying this result, we find that (284) is upper bounded by

$$\begin{aligned} &I\left(X_n; Y_n|Z_n\right) + I\left(W, M_1, Y_f^n; Z_n|M_2, Y^{n-1}, Z^{n-1}\right) \\ &+ I\left(W, M_1, Y_f^n; Y^{n-1}, Z^{n-1}, M_2\right) - I\left(W, M_1, Y_f^n; Z^n\right) \end{aligned} \tag{292}$$

$$\begin{aligned} =&I\left(X_n; Y_n|Z_n\right) + I\left(W, M_1, Y_f^n; Z_n|M_2, Y^{n-1}, Z^{n-1}\right) \\ &+ I\left(W, M_1, Y_f^n; Y^{n-1}, Z^{n-1}, M_2\right) - I\left(W, M_1, Y_f^n; Z^{n-1}\right) \\ &- I\left(W, M_1, Y_f^n; Z_n|Z^{n-1}\right) \end{aligned} \tag{293}$$

We next bound the second term and the last term in (293) together, as shown below:

$$I\left(W, M_1, Y_f^n; Z_n|M_2, Y^{n-1}, Z^{n-1}\right) - I\left(W, M_1, Y_f^n; Z_n|Z^{n-1}\right) \tag{294}$$

$$\begin{aligned} =&h\left(Z_n|M_2, Y^{n-1}, Z^{n-1}\right) - h\left(Z_n|W, M_1, Y_f^n, M_2, Y^{n-1}, Z^{n-1}\right) \\ &- h\left(Z_n|Z^{n-1}\right) + h\left(Z_n|Z^{n-1}, W, M_1, Y_f^n\right) \end{aligned} \tag{295}$$

$$\begin{aligned} =&-I\left(Z_n; M_2, Y^{n-1}|Z^{n-1}\right) \\ &- h\left(Z_n|W, M_1, Y_f^n, M_2, Y^{n-1}, Z^{n-1}\right) + h\left(Z_n|Z^{n-1}, W, M_1, Y_f^n\right) \end{aligned} \tag{296}$$

$$\leq -h\left(Z_n|W, M_1, Y_f^n, M_2, Y^{n-1}, Z^{n-1}\right) + h\left(Z_n|Z^{n-1}, W, M_1, Y_f^n\right) \tag{297}$$

$$= -h\left(Z_n|X_n, W, M_1, Y_f^n, M_2, Y^{n-1}, Z^{n-1}\right) + h\left(Z_n|X_n, Z^{n-1}, W, M_1, Y_f^n\right) \tag{298}$$

$$= -h\left(Z_n|X_n\right) + h\left(Z_n|X_n\right) \tag{299}$$

$$= 0 \tag{300}$$

In (298), we use the fact that $X_n$ is a deterministic function of $W, M_1, Y_f^n$. In (299), we use the fact that $Z_n - X_n - \{W, M_1, Y_f^n, M_2, Y^{n-1}, Z^{n-1}\}$ is a Markov chain and $Z_n - X_n -$ $\{Z^{n-1}, W, M_1, Y_f^n\}$ is a Markov chain. Both are a consequence of the fact that the channel is memoryless and the encoding functions (109) and (110) are causal.

Applying this result to (293), we find it is upper bounded by:

$$I\left(X_n; Y_n|Z_n\right) + I\left(W, M_1, Y_f^n; Y^{n-1}, Z^{n-1}, M_2\right) - I\left(W, M_1, Y_f^n; Z^{n-1}\right) \tag{301}$$

The second term in (301) can be combined with the last term in (301) and expressed as:

$$I\left(W, M_1, Y_f^n; Y^{n-1}, M_2|Z^{n-1}\right) \tag{302}$$

$$= I\left(W, M_1, Y_f^{n-1}; Y^{n-1}, M_2|Z^{n-1}\right) + I\left(Y_{f,n}; Y^{n-1}, M_2|W, M_1, Y_f^{n-1}, Z^{n-1}\right) \tag{303}$$

The last term in (303) can be upper bounded as:

$$I\left(Y_{f,n}; Y^{n-1}, M_2|W, M_1, Y_f^{n-1}, Z^{n-1}\right) \tag{304}$$

$$\leq h\left(Y_{f,n}\right) - h\left(Y_{f,n}|W, M_1, Y_f^{n-1}, Z^{n-1}, Y^{n-1}, M_2\right) \tag{305}$$

$$= h\left(Y_{f,n}\right) - h\left(Y_{f,n}|X_{f,n}, W, M_1, Y_f^{n-1}, Z^{n-1}, Y^{n-1}, M_2\right) \tag{306}$$

$$= h\left(Y_{f,n}\right) - h\left(Y_{f,n}|X_{f,n}\right) \tag{307}$$

$$= I\left(X_{f,n}; Y_{f,n}\right) \tag{308}$$

In (306), we use the fact that $X_{f,n}$ is a deterministic function of $Y^{n-1}, M_2$. (307) follows because $Y_{f,n} - X_{f,n} - \{W, M_1, Y_f^{n-1}, Z^{n-1}, Y^{n-1}, M_2\}$ is a Markov chain.

Applying this result to (301), we find (301) can be upper bounded as:

$$I\left(X_n; Y_n|Z_n\right) + I\left(X_{f,n}; Y_{f,n}\right) + I\left(W, M_1, Y_f^{n-1}; Y^{n-1}, M_2|Z^{n-1}\right) \tag{309}$$

Hence we have shown

$$H(W) - n\varepsilon \leq I\left(W, M_1, Y_f^n; Y^n, M_2|Z^n\right)$$
$$\leq I\left(X_n; Y_n|Z_n\right) + I\left(X_{f,n}; Y_{f,n}\right) + I\left(W, M_1, Y_f^{n-1}; Y^{n-1}, M_2|Z^{n-1}\right) \tag{310}$$

Applying this result repeatedly for $n-1, n-2, ..., 1$, we have

$$\frac{1}{n}H(W) - \varepsilon \leq \frac{1}{n}\sum_{i=1}^{n}(I(X_i; Y_i|Z_i) + I(X_{f,i}; Y_{f,i})) \tag{311}$$

Let us define $Q$ as a random variable that is uniformly distributed over $\{1, 2, ..., n\}$. Further, define $X = X_Q, Y = Y_Q, Z = Z_Q, X_f = X_{f,Q}, Y_f = Y_{f,Q}$. Then, the right hand side of (311) can be expressed as

$$I(X; Y|Z, Q) + I(X_f; Y_f|Q) \tag{312}$$

$$\leq h\left(Y|Z\right) - h\left(Y|X,Z,Q\right) + h\left(Y_f\right) - h\left(Y_f|X_f,Q\right) \tag{313}$$

$$= h\left(Y|Z\right) - h\left(Y|X,Z\right) + h\left(Y_f\right) - h\left(Y_f|X_f\right) \tag{314}$$

$$= I\left(X;Y|Z\right) + I\left(X_f;Y_f\right) \tag{315}$$

Applying this result in (303) and letting $n \to \infty$, we obtain the upper bound in the theorem.